\documentclass[reprint,superscriptaddress,
amsmath,amssymb,
pra,
]{revtex4-1}
\usepackage[utf8]{inputenc}

\usepackage{graphicx}
\usepackage{dcolumn}
\usepackage{bm}
\usepackage{hyperref}
\usepackage{ulem}
\usepackage{xcolor}
\usepackage{amsmath, amssymb}
\usepackage[T1]{fontenc}

\renewcommand{\vec}[1]{{\boldsymbol{#1}}} 
\newcommand{\ave}[1]{\left< #1 \right>}
\usepackage{braket}

\begin{document}
	
	\preprint{APS/123-QED}

	
	\title{Towards a QMC-based density functional including finite-range effects: \\ excitation modes of a $^{39}$K quantum droplet}
	
	\author{V. Cikojevi\'c}
	\affiliation{University of Split, Faculty of Science, Ru\dj era Bo\v{s}kovi\'ca 33, HR-21000 Split, Croatia}
	\affiliation{Departament de F\'{\i}sica, Universitat Polit\`ecnica de Catalunya, Campus Nord B4-B5, E-08034 Barcelona, Spain}
	
	\author{L. Vranje\v{s} Marki\'{c}}
	\affiliation{University of Split, Faculty of Science, Ru\dj era Bo\v{s}kovi\'ca 33, HR-21000 Split, Croatia}
	
	\author{M. Pi}
	\affiliation{Departament FQA, Facultat de F\'{\i}sica, Universitat de Barcelona, Diagonal 645, 08028 Barcelona, Spain}
	\affiliation{Institute of Nanoscience and Nanotechnology (IN2UB), Universitat de Barcelona, 08028 Barcelona, Spain}
	
	\author{M. Barranco} 
	\affiliation{Departament FQA, Facultat de F\'{\i}sica, Universitat de Barcelona, Diagonal 645, 08028 Barcelona, Spain}
	\affiliation{Institute of Nanoscience and Nanotechnology (IN2UB), Universitat de Barcelona, 08028 Barcelona, Spain}
	
	\author{J. Boronat} 
	\affiliation{Departament de F\'{\i}sica, Universitat Polit\`ecnica de Catalunya, Campus Nord B4-B5, E-08034 Barcelona, Spain}

	\date{\today}
	
	\begin{abstract}
		Some discrepancies between experimental results on quantum droplets made of a mixture of $^{39}$K atoms  in different hyperfine states 
		and their analysis within extended Gross-Pitaevskii 
		theory (which incorporates beyond mean-field corrections) have been recently solved by introducing finite-range effects into the theory. 
		Here, we study the influence of these effects on the monopole and quadrupole excitation spectrum of extremely dilute quantum droplets using
		a density functional built from first-principles quantum Monte Carlo calculations, which can be easily introduced in the existing Gross-Pitaevskii 
		numerical solvers. Our results show differences of up to $20\%$ with those obtained within the extended Gross-Pitaevskii theory, likely providing another 
		way to  observe finite-range effects in  mixed quantum droplets by measuring their lowest excitation frequencies.
	\end{abstract}
	
	\pacs{}

	\maketitle
	
	\section{Introduction}

	Ultracold gases serve as a unique platform for understanding quantum many-body physics~\cite{bloch2008many}. This notoriously hard problem is often reduced to the effective single-particle picture when the interactions are very weak and the density is 
	very low~\cite{pethick2008bose,pitaevskii2016bose}. 
	Because of its simplicity and predictive power, the mean field approach has become a standard (or a first starting point)  to study the properties of ultracold gases.	
	
	The accuracy of mean-field theories to address dilute quantum gases is expectable, as nearly all experiments  are 
	performed at very low values of the gas parameter $\rho a^3$,  $\rho$ being the atom number density and $a$  the s-wave scattering length 
	describing the interparticle interactions. 
	This allows for a perturbative approach \`a la Bogoliubov~\cite{bogoliubov1947theory}, where static and dynamic properties are well described by 
	the Gross-Pitaevskii equation. However, as the density and/or the interaction strength increases, the system becomes more correlated and out 
	of the range of applicability of perturbation theories. It is a priori difficult to know when the perturbative approach is no longer valid. Thus, 
	it is essential to supplement the theory with developments \cite{cikojevic2018ultradilute,cikojevic2019universality,parisi2020quantum,parisi2019liquid,staudinger2018self,petrov2016ultradilute,bombin2017dipolar,ancilotto2018self, hu2020consistent,hu2020microscopic,ota2020beyond} 
	aiming at verifying the range of applicability of the mean-field approach and disclosing  the role played by higher-order effects.
	
	A promising system for investigating  quantum many-body effects, going beyond mean-field theory, is the self-bound Bose-Bose mixture 
	first proposed by Petrov \cite{petrov2015quantum}. In this mixture, with repulsive intraspecies and attractive interspecies short-range interactions,
	the unstable attractive mean-field energy is balanced out by a repulsive beyond mean-field term (the Lee-Huang-Yang (LHY) term) 
	\cite{lee1957eigenvalues}, resulting in a liquid droplet resembling the well-known  $^4$He droplets~\cite{barranco2006helium}, but with a
	far smaller density. So far, a Bose-Bose droplet state has been observed in a mixture of two $^{39}$K hyperfine states \cite{cabrera2018quantum,semeghini2018self,cheiney2018bright}, and in an heterogeneous mixture of $^{41}$K-$^{87}$Rb 
	atoms \cite{derrico2019observation}. 
	
	In the first experimental observation~\cite{cabrera2018quantum},  discernible differences were observed between the experiment and the results of the mean-field (MF) theory extended with an LHY term. Quite recently, it has been reported~\cite{cikojevic2020finite} that the agreement 
	between theory and experiment improves notably when finite-range effects are properly taken into account. For the particular mixture 
	of two hyperfine states of $^{39}$K atoms, we know two scattering parameters in each of the interaction channels \cite{tanzi2018feshbach}, 
	the s-wave scattering length $a$ and the effective range $r^{\rm eff}$, which are the first two coefficients in the expansion of the s-wave phase 
	shift in the scattering between two atoms \cite{newton2013scattering}
	\begin{equation}
	k \cot \delta (k) = -\dfrac{1}{a} + \dfrac{1}{2} r^{\rm eff} k^2 + \mathcal{O}(k^4).
	\end{equation}
	The non-zero (in fact quite large) effective ranges open a promising new regime in quantum mixtures which go beyond the 
	usual mean-field theory corrected with the LHY term (MF+LHY) \cite{tononi2019zero,tononi2018condensation,salasnich2017nonuniversal}.  
	A  large effective range means that the interaction between atoms is far from the contact Dirac $\delta$-interaction  usually employed for 
	dilute Bose gases. 
	
	In a previous work~\cite{cikojevic2020finite}, some of us have performed diffusion Monte Carlo (DMC) calculations 
	\cite{boronat1994monte,giorgini1999ground} using model potentials that reproduce both scattering parameters, obtaining the equation of state 
	for a $^{39}$K mixture in the homogeneous liquid phase.  We concluded that  one could
	reproduce the  critical atom number determined  in the experiment \cite{cabrera2018quantum} only for the model potentials which incorporate 
	the correct effective range. This critical number is a static property  of the  quantum droplet at equilibrium. Besides a good knowledge of the
	equilibrium properties of  a quantum many-body system, determining the excitation spectrum is essential to unveil its microscopic structure. 
	
	In the present work, we present a study of the monopole and quadrupole excitation spectrum of a $^{39}$K quantum droplet using 
	the QMC functional introduced in Ref.~\cite{cikojevic2020finite}, which correctly describes the inner part of large drops, constituting 
	an extension to the MF+LHY theory.    The excitation spectrum of these droplets has already been calculated within the MF+LHY 
	approach \cite{petrov2015quantum,jorgensen2018dilute}. Our goal is to make visible  the appearance of any  beyond-LHY effect arising from 
	the inclusion of the effective range in the interaction potentials.

	This paper is organized as follows. We build in Sec. \ref{sec:functional_form} the QMC density functional, in the local density approximation (LDA), and
	compare it with the MF+LHY approach, which can be expressed in a similar form. In Sec. \ref{sec:method}, we give details on the application of the density 
	functional method, static and dynamic,  to the obtainment of the ground state and excitation spectrum of quantum droplets.
	In Sec. \ref{sec:results}, we report the results of the monopole and quadrupole frequencies obtained with the QMC functional and compare them 
	with the MF+LHY predictions. Finally, a summary and outlook are presented in  Sec. \ref{sec:summary}.

	\section{\label{sec:functional_form} The QMC Density Functional}
	
	We shall consider  $^{39}$K mixtures at   
	the optimal relative atom concentration yielded by the mean-field theory, namely $N_1 / N_2 = \sqrt{a_{22} / a_{11}}$ \cite{petrov2015quantum}. 
	For these mixtures, we have  shown that the energy per atom in the QMC approach can be  accurately written as \cite{cikojevic2020finite}        
	\begin{equation}
	\dfrac{E}{N} = \alpha \rho + \beta \rho^{\gamma} \ ,
	\label{eos}
	\end{equation}
	where $\rho$ is the total atom number density. The parameters $\alpha$, $\beta$, and $\gamma$ 
	have been determined by fits to the DMC results for the model potentials satisfying the s-wave scattering length and effective range, given in Table \ref{table:scattering_parameters}. Parameters appearing in Eq. (\ref{eos}) are collected in Table \ref{table:eos_params}, for three values of the magnetic field ($B$).
	The QMC approach does not yield a universal  expression for $E/N$,
	as it depends on the value of the applied $B$. 
	For the optimal concentration,  the MF+LHY energy per particle can be cast in a similar expression 
	\begin{equation}
	\label{eq:mflhy_eos}
	\dfrac{E/N}{|E_0|/ N} = -3\left(\dfrac{\rho}{\rho_0}\right) + 2 
	\left(\dfrac{\rho}{\rho_0}\right)^{3/2} \ ,
	\end{equation}
	where $E_0/N$ and $\rho_0$ are the energy per atom and atom density at equilibrium,
	\begin{equation}
	\label{eq:en_0}
	E_0 / N = \dfrac{25 \pi^2 \hbar^2 |a_{12} + \sqrt{a_{11} 
			a_{22}}|^3}{768m a_{22} a_{11} \left(\sqrt{a_{11}} + \sqrt{a_{22}}\right)^6} 
	\end{equation} 
	and
	\begin{equation}
	\label{eq:rho_0}
	\rho_0  = \dfrac{25 \pi }{1024 a_{11}^3} \dfrac{\left(a_{12}/a_{11} + 
		\sqrt{a_{22}/a_{11}}\right)^2}{\left(a_{22}/a_{11}\right)^{3/2}\left(1+\sqrt{a_{
				22}/a_{11}}\right)^4} \ .
	\end{equation}
	In Eqs. (\ref{eq:en_0}) and (\ref{eq:rho_0}), $m$ is the mass of a $^{39}$K atom and $a_{ij}$ are the three different s-wave scattering lengths. MF+LHY theory 
	is thus universal if it is expressed in terms of $\rho_0$ and $E_0$. According to this theory,  the droplet properties do not change separately 
	on $N$ and $a_{ij}$ but rather combined through
	\begin{equation}
	\label{eq:ntilde_def}
	\dfrac{N}{\tilde{N}} = \dfrac{3\sqrt{6}}{5\pi^2} \dfrac{\left(1+\sqrt{a_{22}/a_{11}}\right)^5}
	{\left|a_{12} / a_{11} + \sqrt{a_{22} / a_{11}}\right|^{5/2}} \ ,
	\end{equation}
	where $\tilde{N}$ is a dimensionless parameter~\cite{petrov2015quantum}. Additionally, the healing length corresponding to the mixture is
	\begin{equation}
	\label{eq:xi_mflhy}
	\dfrac{\xi}{a_{11}} = 
	\dfrac{8\sqrt{6}}{5\pi} \sqrt{\dfrac{a_{22}}{a_{11}}} \dfrac{(1 + \sqrt{a_{22} / a_{11}})^3}{\left|a_{12}/a_{11} + \sqrt{a_{22} / a_{11}}\right|^{3/2}} \ .
	\end{equation}

	\begin{table}[tb]
		\caption{Scattering parameters, i.e. s-wave scattering length $a$ and  effective range $r^{\rm eff}$ (in units of Bohr 
			radius $a_0$) as a function of $B$ \cite{roy2013test}.  }
		\label{table:scattering_parameters}
		\begin{tabular}{ c | c | c | c | c | c | c }
			\hline
			$B ({\rm G})$ & $a_{11} (a_0)$ & $r_{11}^{\rm eff}(a_0)$ & $a_{22} (a_0)$ & $r_{22}^{\rm eff}(a_0)$ & $a_{12} (a_0)$ & $r_{12}^{\rm eff}(a_0)$ \\
			\hline 
			56.230 & 63.648 & -1158.872  & 34.587  & 578.412 & -53.435 & 1021.186  \\
			56.453 & 70.119 & -1150.858  & 34.136  & 599.143 & -53.333 & 1023.351  \\
			56.639 & 76.448 & -1142.642  & 33.767  & 616.806 & -53.247 & 1025.593  \\
		\end{tabular}
	\end{table}
	\begin{table}[tb]
		\caption{Parameters of the QMC energy per atom calculated at several magnetic fields $B$, 
			assuming $\rho_1/\rho_2 = \sqrt{a_{22} / a_{11}}$, satisfying the s-wave scattering length $a$ and effective range $r^{\rm eff}$ 
			given in Table \ref{table:scattering_parameters}. $\alpha$ is  in $\hbar^2 a_{11}^2 / (2m)$ units, $\beta$ is in 
			$\hbar^2 a_{11}^{3\gamma - 2} / (2m)$ units,
			$m$ being the mass of a $^{39}$K atom, and $\gamma$ is dimensionless.}
		\label{table:eos_params}
		\begin{tabular}{c  | c | c | c }
			\hline
			$B({\rm G})$  &  $\alpha$ & $\beta$ & $\gamma$ \\ \hline
			56.230  & -0.812     &  5.974  &  1.276    \\ 
			56.453    & -0.423        &  8.550    &  1.373 \\ 
			56.639    & -0.203       &  12.152    & 1.440 \\  
		\end{tabular}
	\end{table}

	\begin{figure}
		\centering
		\includegraphics[width=\linewidth]{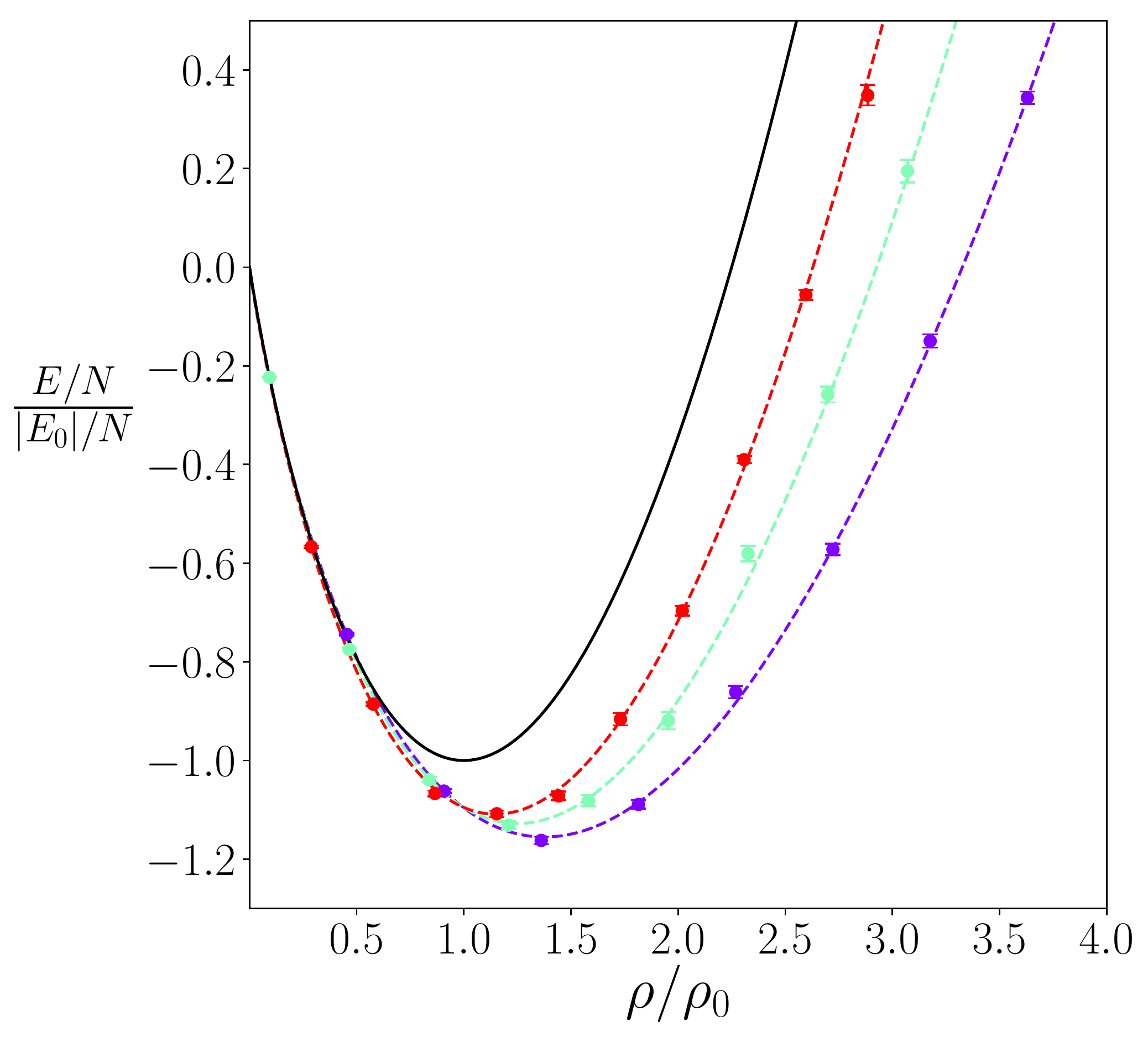}
		\caption{DMC energy per particle as a function of the density. From bottom (blue dots) to top (red dots), the results correspond to
			magnetic fields  $B$=56.23, 56.453 and 56.639 G.
			Calculations were performed for the mean-field optimal  ratio $\rho_2 / \rho_1 = \sqrt{{a_{11} / a_{22}}}$. The energy per 
			atom  and atom density 
			are normalized to the  $|E_0|/N$ and $\rho_0$ MF+LHY values obtained  from Eqs. (\ref{eq:en_0}) and (\ref{eq:rho_0}), respectively. 
			The dashed lines are fits in the form $E/N = \alpha \rho + \beta \rho^\gamma$. The black solid line corresponds to the MF+LHY theory, 
			Eq. (\ref{eq:mflhy_eos}).} 
		\label{fig:eoscombineddiffbuniversal}
	\end{figure}
	
	The energy per atom Eq. (\ref{eos}) allows one to readily introduce, within LDA,  a density functional whose interacting part is 
	\begin{equation}
	\mathcal{E}_{\rm int} = \rho \frac{E}{N} = \alpha \rho^2 + \beta\rho^{\gamma + 1} \ .
	\end{equation}
	A similar expression     holds in the MF+LHY approach.    
	In the homogeneous phase, one may easily obtain the pressure 
	\begin{equation}
	p(\rho) = \rho^2 \dfrac{\partial}{\partial \rho}\left(\dfrac{E}{N}\right) = \alpha \rho^2 + \beta \gamma \rho^{\gamma+1}
	\end{equation}
	and incompressibility
	\begin{equation}
	\kappa(\rho) =\rho \frac{\partial p}{\partial \rho},
	\end{equation}
	which can be written as 
	\begin{equation}
	\kappa(\rho) =\rho^2 \frac{\partial^2 \mathcal{E}_{\rm int}}{\partial \rho^2}=
	\rho^2\left\{ 2 \frac{\partial}{\partial \rho} \left( \frac{E}{N}\right) + \rho \frac{\partial^2}{\partial \rho^2} \left(\frac{E}{N}\right) \right\}.
	\label{eq:K}
	\end{equation}
	
\begin{figure}
		\centering
		\includegraphics[width=\linewidth]{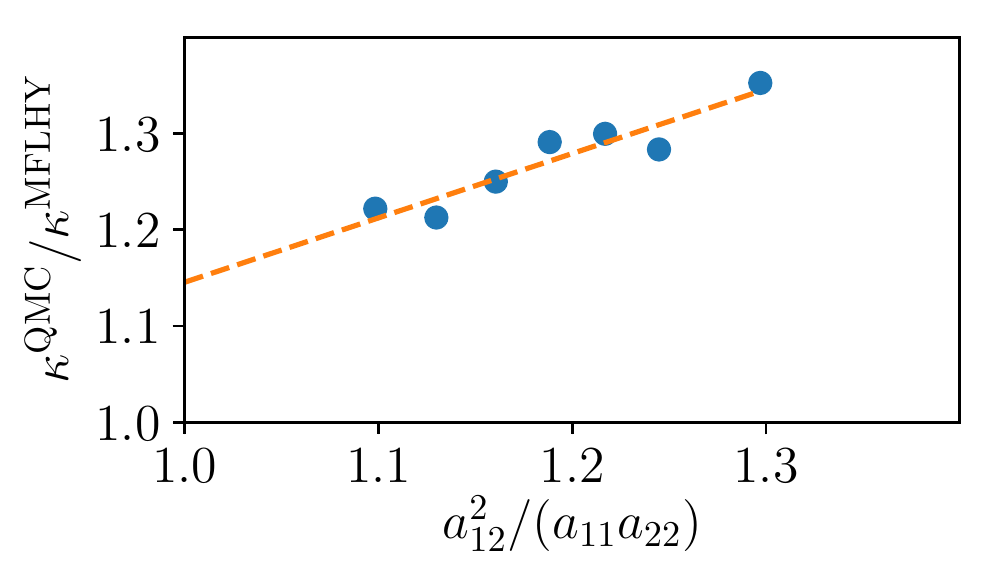}
		\caption{DMC  over MF+LHY incompressibility  ratio at equilibrium for the magnetic fields  considered in  Ref.~\cite{cikojevic2020finite}. The dashed line is a linear fit to the points.
		}
		\label{fig:KQMC-o-MF}
	\end{figure}

\begin{figure}
		\centering
		\includegraphics[width=\linewidth]{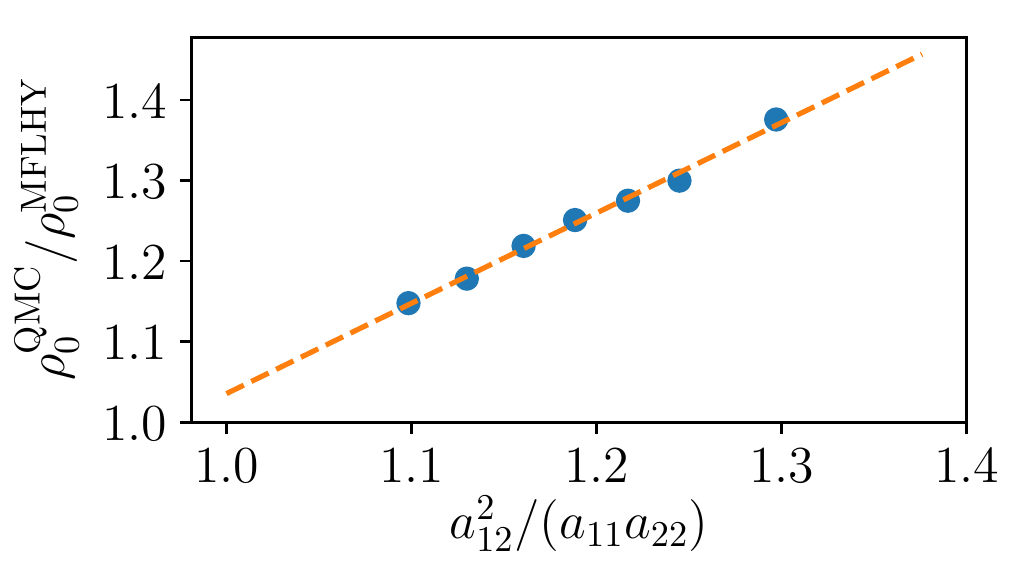}
		\caption{DMC  over MF+LHY equilibrium density ratio for the magnetic fields  considered in  Ref.~\cite{cikojevic2020finite}. The dashed line is a linear fit to the points.
		}
		\label{fig:rhoQMC-o-MF}
	\end{figure}

	Figure  \ref{fig:eoscombineddiffbuniversal} shows the  DMC energy per atom as a function of the density for selected values of the magnetic field, together
	with the result for the  MF+LHY theory. It is worth noticing the rather different equations of state yielded by the QMC functional and MF+LHY approaches. 
	The QMC approach yields a substantially larger equilibrium density and more binding. The QMC incompressibility is also larger,  as can be 
	seen in Fig. \ref{fig:KQMC-o-MF};
	at first sight, this seems to be in contradiction with the results in Fig.  \ref{fig:eoscombineddiffbuniversal}, which clearly indicate that the curvature of the 
	$E/N$ vs $\rho$ curve at equilibrium ($\partial (E/N)/\partial \rho=0$ point) is smaller for the QMC functionals than for the MF+LHY approach. 
	However, this is compensated by the larger QMC value of the atom density at equilibrium, see Eq. (\ref{eq:K}) and  Fig. \ref{fig:rhoQMC-o-MF},
	where we  show the ratio of QMC and MF+LHY equilibrium densities. Besides its importance for a quantitative description of the
	monopole droplet oscillations addressed here, inaccurate incompressibility may affect the description of processes  where the liquid-like
	properties of quantum droplets play a substantial role, as {\it e.g.} droplet-droplet collisions  \cite{ferioli2019collisions}. 
	
	\begin{table}[tb]
		\caption{Surface tension of a $^{39}$K Bose-Bose mixture at the MF+LHY optimal mixture composition 
			in $10^{-8} \times \hbar^2 / (m a_{11}^4)$ units.}
		\label{table:surface_tension}
		\begin{tabular}{c  | c | c  }
			\hline
			$B({\rm G})$  &  $\sigma_{\rm MF+LHY}$ & $\sigma_{\rm QMC}$ \\ \hline
			56.230  &  35.1   &  48.8    \\ 
			56.453    &    9.31     &  12.2 \\ 
			56.639    &    1.21    & 1.46     \\ 
		\end{tabular}
	\end{table}

	Another fundamental property of  the liquid is the surface tension $\sigma$ of the free-surface. Remarkably, for simple functionals as the 
	QMC and MF+LHY ones discussed in this work, its value  can be obtained by  simple quadrature \cite{stringari1985surface}         
	\begin{equation}
	\label{eq:surface_tension_formula}
	\sigma = 2 \int_{0}^{\rho_0} d\rho \left[\left(\dfrac{\hbar^2}{8m} \right) \left(\alpha \rho + \beta \rho^{\gamma} - \mu\right)\right]^{1/2},
	\end{equation}
	where $\mu$ is the chemical potential evaluated at the equilibrium density. The surface tension of several QMC functionals, i.e. functionals corresponding to different magnetic fields, is given in Table \ref{table:surface_tension}. As can be seen, QMC functionals yield consistently 
	higher values of the surface tension than the MF+LHY approach.	
	Within MF+LHY, the surface tension can be written in terms of the equilibrium density (\ref{eq:rho_0}) and healing length (\ref{eq:xi_mflhy}),
	$\sigma_{\rm MF+LHY} = 3(1 + \sqrt{3}) \rho_0 \hbar^2 / (35 m \xi )$ \cite{petrov2015quantum}.

	\section{\label{sec:method} The LDA-DFT approach}
	
	\subsection{Statics}
	
	Once $\mathcal{E}_{\rm int}[\rho]$ has been obtained, we have used density functional theory (DFT) to address the static and dynamic 
	properties of $^{39}$K droplets similarly as for superfluid $^4$He droplets \cite{ancilotto2017density}.    
	Within DFT, the energy of the quantum droplet  at the optimal composition mixture is written as a functional of the atom density $\rho({\mathbf r})$ as
	\begin{equation}
	E[\rho] = T[\rho] + E_c[\rho] =
	\frac{\hbar^2}{2m} \int d {\mathbf r} |\nabla \Psi({\mathbf r})|^2 +  \int d{\mathbf r} \,{\cal E}_{\rm int}[\rho],
	\label{eq1}
	\end{equation}
	where the first term  is the kinetic energy, and the effective wavefunction $\Psi({\mathbf r})$ of the droplet  is  related to the
	atom density as $\rho({\mathbf r})= |\Psi({\mathbf r})|^2$.  
	The equilibrium configuration is obtained by solving the Euler-Lagrange  equation arising 
	from the functional minimization of Eq. (\ref{eq1})
	\begin{equation}
	\left\{-\frac{\hbar^2}{2m} \nabla^2 + \frac{\partial {\cal E}_{int}}{\partial \rho}  \right\}\Psi 
	\equiv {\cal H}[\rho] \,\Psi  = \mu \Psi,
	\label{eq3}
	\end{equation}
	where $\mu$ is the chemical potential corresponding to the number of $^{39}$K atoms in the droplet, $N = \int d{\bf r}|\Psi({\bf r})|^2$. 
	
	The time-dependent  version of  Eq. (\ref{eq3}) is obtained minimizing the action and adopts the form
	\begin{equation}
	i \hbar \frac{\partial}{\partial t}  \Psi({\mathbf r},t) = {\cal H}[\rho] \,\Psi({\mathbf r},t).
	\label{eq4}
	\end{equation}
	We have implemented a three-dimensional numerical solver based 
	on the Trotter decomposition of the time-evolution 
	operator with second-order accuracy  in the time-step $\Delta t$ ~\cite{chin2009any}
	\begin{equation}
	e^{-i {\cal H} \Delta t}  = e^{-i\Delta t V(\vec{R}') / 2}  e^{-i\Delta     t     K } e^{-i\Delta t V(\vec{R})/ 2} + \mathcal{O}(\Delta t^3) \ ,
	\end{equation}
	with $K$ and $V$ being the kinetic and interaction terms in Eq. (\ref{eq3}). Within this scheme, it is possible to obtain both the ground 
	state and the dynamical evolution. Indeed, reformulating the problem via a Wick rotation $t = -i\tau$, the propagation of a wavefunction 
	in imaginary time $\tau$  leads to the 
	ground-state equilibrium solution.
	
	\begin{figure}
		\centering
		\includegraphics[width=\linewidth]{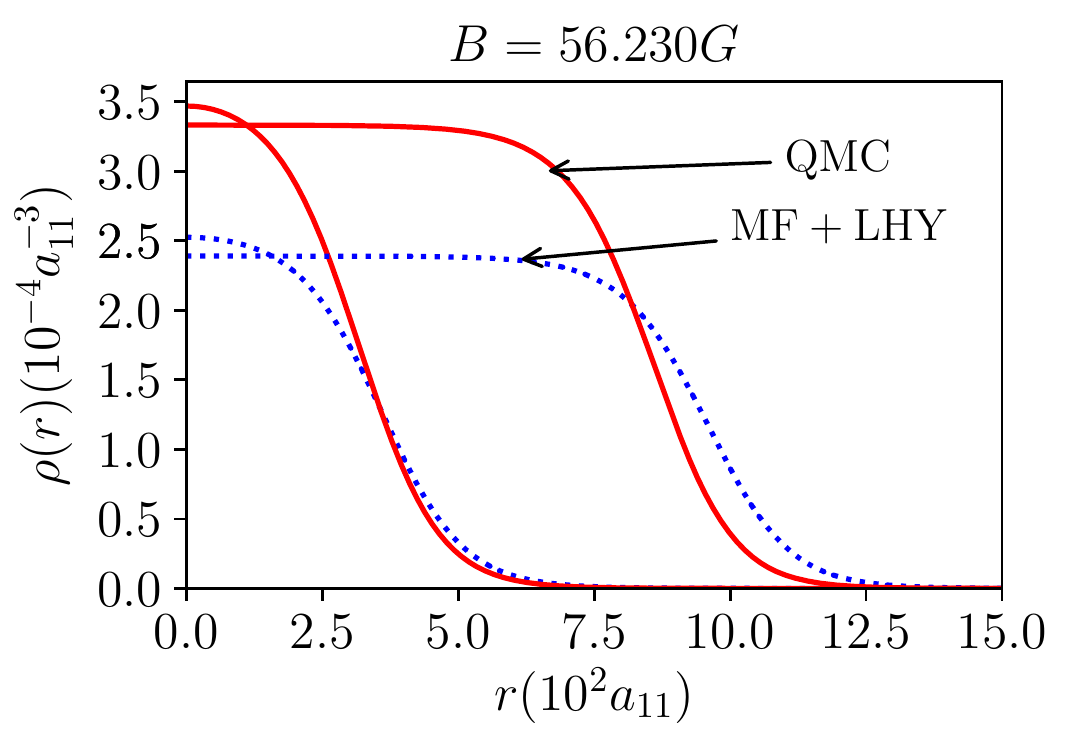}
		\caption{
			Density profiles of two $^{39}$K quantum droplets corresponding to a small  $(\tilde{N} -\tilde{N}_c)^{1/4}=3$, and to a large droplet
			$(\tilde{N} -\tilde{N}_c)^{1/4}=6$, where $\tilde{N}_c=18.65$ is the critical number below which the droplet becomes unstable within the MF+LHY theory \cite{petrov2015quantum}. Solid lines, QMC functional; dotted lines, MF+LHY approach.
		}
		\label{fig:density-profiles}
	\end{figure}    
	
	Figure \ref{fig:density-profiles} shows the density profile of two droplets, one corresponding to a small gaussian-like droplet and the other to a large saturated one. They have
	been obtained within the QMC ($B=56.230$ G) functional and MF+LHY methods. The sizeable difference between the 
	profiles yielded by both approaches reflects the different value of their equilibrium densities, see Fig. \ref{fig:rhoQMC-o-MF}.
	
	\subsection{Real-time dynamics and excitation spectrum}

	The multipole excitation spectrum of a quantum droplet can be obtained {\it e.g.} by solving the equations
	obtained linearizing Eq. (\ref{eq4}) \cite{dalfovo1999theory,petrov2015quantum,baillie2017collective}. We have used an equivalent method based on 
	the Fourier analysis of the real-time oscillatory response of the droplet to an appropriated external field \cite{stringari1979damping,pi1986time}.    
	The method, which we outline now, bears clear similarities with the experimental procedure to access to some excited states of confined 
	Bose-Einstein condensates (BEC) \cite{jin1996collective,altmeyer2007precision}. 
	
	A droplet at the equilibrium, whose ground-state effective wavefunction $\Psi({\mathbf r})$ is obtained by solving the DFT Eq. (\ref{eq3}), is displaced from it by the action of a static  external  one-body field $Q$ whose intensity is controlled by a parameter $\lambda$. The new equilibrium wavefunction $\Psi'({\mathbf r})$ is
	determined by solving Eq. (\ref{eq3}) for the constrained Hamiltonian ${\cal H}'$
	\begin{equation}
	{\cal H} \rightarrow {\cal H}' = {\cal H} + \lambda Q .
	\label{eq16}
	\end{equation}
	If $\lambda$ is  small enough so that $\lambda Q$ is a perturbation and linear response theory applies, switching off $Q$  and letting 
	$\Psi'({\mathbf r})$ evolve in time according to Eq. (\ref{eq4}), $\langle Q(t) \rangle$ will oscillate around the equilibrium value
	$Q_{eq}=\langle \Psi({\mathbf r}) | Q | \Psi({\mathbf r})\rangle$.  Fourier analyzing $\langle Q(t) \rangle$, one gets the non-normalized strength function  corresponding to the excitation operator $Q$, which displays peaks at the frequency values corresponding to the excitation modes of the droplet. Specific values of $\lambda$ that we use are in the range from  $\lambda = 10^{-13}$ to $10^{-15}$ for the monopole modes, and $\lambda = 10^{-15}$ to $10^{-17}$ for the quadropole modes, with $\lambda$ being measured in $\hbar^2 / (2m a_{11}^4)$ units, and the smaller values corresponding to larger magnetic fields, i.e. less correlated drops.
	
	\section{\label{sec:results} Results}
	
	We have used as excitation fields the monopole $Q_0$ and quadrupole $Q_2$ operators 
	\begin{eqnarray}
	Q_{\rm 0} & = & \sum_{i}^N  r_i^2 \\
	Q_{\rm 2} & = & \sum_{i}^N \left(r_i^2 - 3z_i^2\right) 
	\end{eqnarray}    
	which allows one to obtain the $\ell=0$ and 2 multipole strengths. The $\ell=0$ case corresponds to pure radial oscillations of the
	droplet and for this reason it is called ``breathing'' mode. In a pure hydrodynamical approach, its frequency is determined by the 
	incompressibility of the liquid  and the radius of the droplet \cite{bohigas1979sum,pitaevskii2016bose}.
	
\begin{figure}
		\centering
		\includegraphics[width=\linewidth]{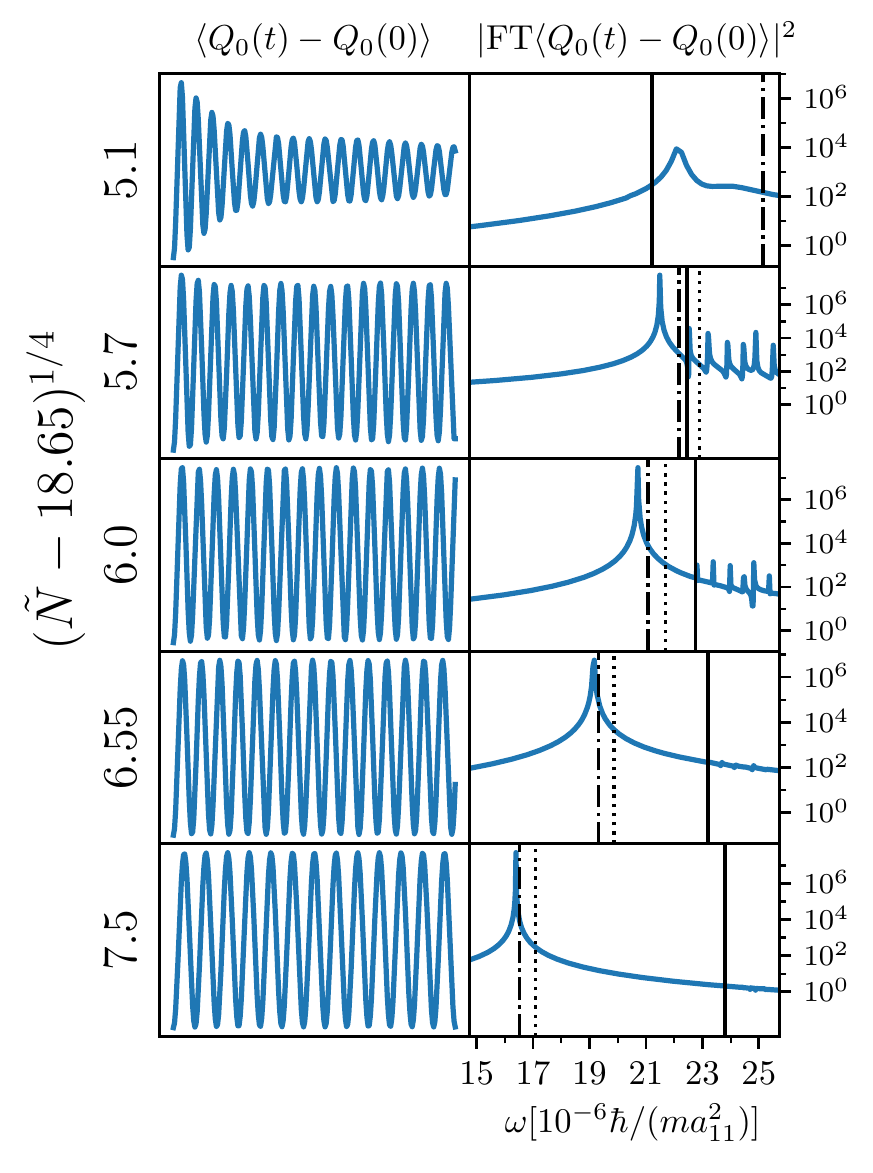}    
		\caption{ 
			Time evolution of the monopole moment $\langle Q_0(t) \rangle$  
			and  strength function (right) for $^{39}$K quantum droplets of different sizes
			obtained  using the QMC functional at $B = 56.230$ G. 
			In the right panels, the vertical solid line corresponds to
			the frequency $|\mu| / \hbar$ corresponding to the atom emission energy $|\mu|$, and the dotted and dash-dotted lines to the $E_3/\hbar$ and $E_1/\hbar$ frequencies, obtained by the sum rules in Eq. (\ref{eq:monopole_E3}) and (\ref{eq:monopole_E1}), respectively.
		}
		\label{fig:monopole}
	\end{figure}

	We have propagated the excited state  $\Psi'({\mathbf r})$ for a very long period of time,  storing  $\langle Q(t) \rangle$ and Fourier analyzing it.
	Fig. \ref{fig:monopole} (left)  shows $\langle Q_0(t) \rangle$ for $^{39}$K quantum droplets of different sizes. We choose the same scale of particle numbers (x-axis) as in Ref. \cite{petrov2015quantum}, as the monopole frequency $\omega_0$ close to the instability point $\tilde{N}_c = 18.65$ is directly proportional to $(\tilde{N} - \tilde{N}_c)^{1/4}$ \cite{petrov2015quantum}. Whereas a harmonic behavior 
	is clearly visible for the largest droplets, as corresponding to a single-mode excitation, for small droplets the radial oscillations are damped 
	and display different oscillatory behaviors (beats), anticipating
	the presence of several modes in the monopole strength, as the Fourier analysis of the signal unveils.
	
	Figure  \ref{fig:monopole} (right) displays the monopole strength function in logarithmic scale as a function of the excitation frequency. The solid vertical line represents the frequency $|\mu| / \hbar$ corresponding to the atom emission threshold, i.e. the absolute value of the atom chemical potential, $|\mu|$. It can be seen that for $(\tilde{N} -18.65)^{1/4} =5.1$  the strength  is in the continuum frequency region above $|\mu|/\hbar$.  
	Hence,  self-bound small $^{39}$K droplets, monopolarly excited, have excited states (resonances) that may decay by atom emission \cite{petrov2015quantum,ferioli2020dynamical}.
	This decay does not imply that the droplet breaks apart; it just loses the energy deposited into it by emitting a number of atoms, in a way  
	similar to the decay of  some states appearing in the atomic nucleus, the so-called ``giant resonances'' \cite{bohigas1979sum}.
	We want to stress that the multipole strength is not normalized, as it depends on the value of the arbitrary small parameter $\lambda$. However, 
	the relative intensity of the peaks for a given droplet is properly accounted for in this approach.
	
\begin{figure}
		\centering
		\includegraphics[width=\linewidth]{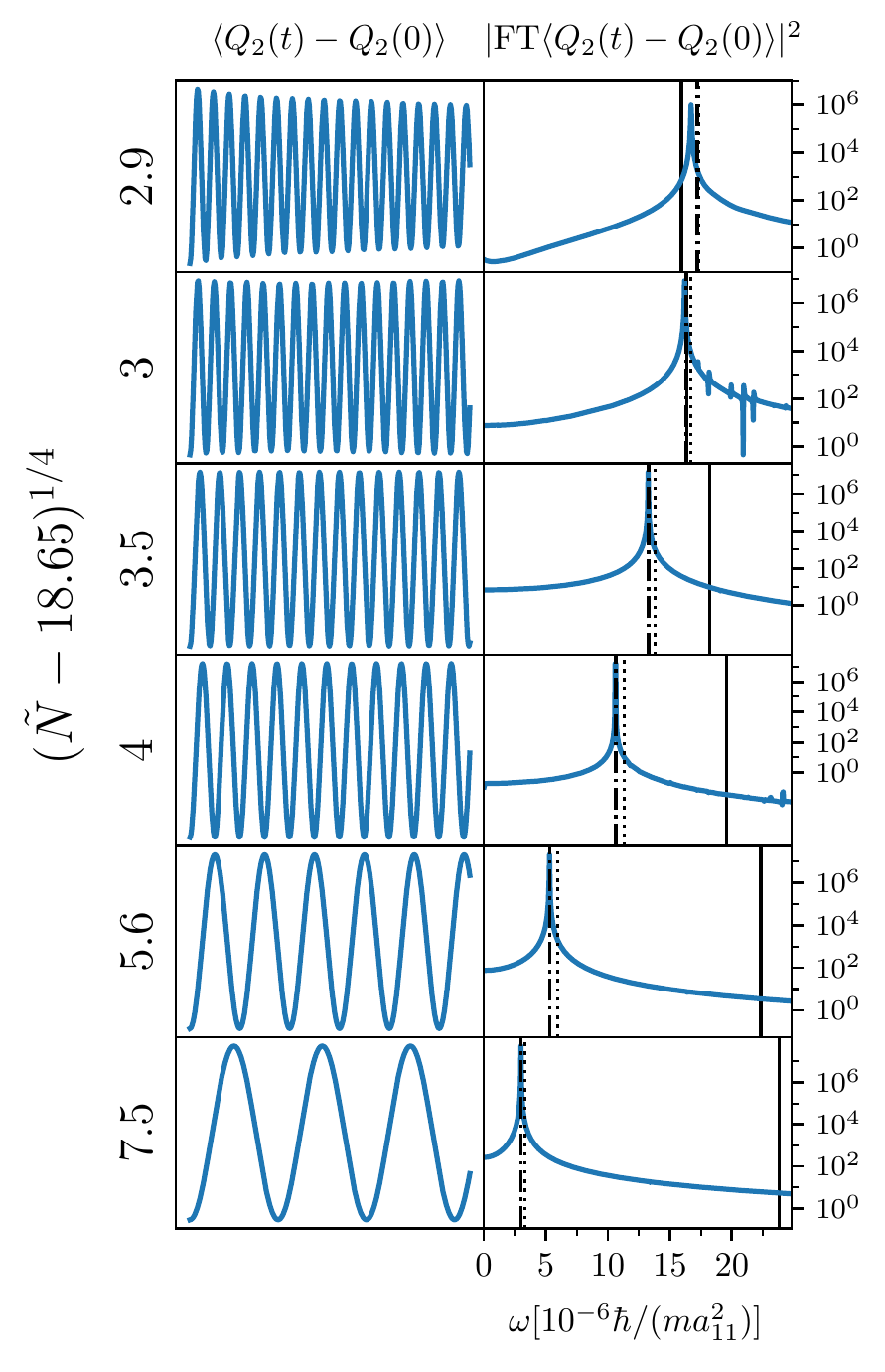}    
		\caption{  
			Time evolution of the quadrupole moment $\langle Q_2(t) \rangle$  
			and  strength function (right) for $^{39}$K quantum droplets of different sizes
			obtained  using the QMC functional at $B = 56.230$ G. 
			In the right panels, the vertical solid line corresponds to
			the frequency $|\mu| /\hbar$ corresponding to the atom emission energy $|\mu|$, and the dotted and dash-dotted lines to the $E_3/\hbar$ and $E_1/\hbar$ frequencies,  obtained by the sum rules in Eq. (\ref{eq:quadropole_E3}) and (\ref{eq:quadropole_E1}), respectively.			
		}
		\label{fig:quadrupole}
	\end{figure}

	A similar analysis for the quadrupole mode is  presented in Fig. \ref{fig:quadrupole}.
	In this case, we have found a more harmonic behavior for $\langle Q_2(t) \rangle$, 
	and therefore the quadrupole strength function  is dominated by one single peak.
	
	Figures  \ref{fig:monopole} and \ref{fig:quadrupole} show an interesting evolution of the strength function from the continuum to the discrete
	part of the frequency spectrum as the number of atoms in the droplet increases. For small $N$ values, but still corresponding to self-bound
	quantum droplets, the spectrum is dominated by  a broad resonance that may  decay by atom emission. The   
	$\langle Q (t) \rangle$ oscillations are damped, and when several resonances	are present (monopole case), distinct beats appear 
	in the oscillations.
	
	This remarkable evolution of the monopole and quadrupole  spectrum  has also been found for $^3$He 
	and $^4$He droplets \cite{serra1991collective,barranco1994response}. In the $^4$He case, it has been  experimentally confirmed by detecting 
	``magic'' atom numbers in the size distribution of $^4$He droplets which correspond to especially stable droplets  \cite{bruhl2004diffraction}. The
	magic numbers occur at the threshold sizes for which the excitation modes of the droplet, as calculated by the diffusion Monte Carlo method, 
	are stabilized when they pass below the atom emission energy. This constituted the  
	first experimental confirmation for the energy levels of $^4$He droplets. On the other hand,  in  confined BECs, the energy of the breathing mode is
	obtained by direct analysis of the radial oscillations of the atom cloud \cite{pitaevskii2016bose}.

\begin{figure}
		\centering
		\includegraphics[width=\linewidth]{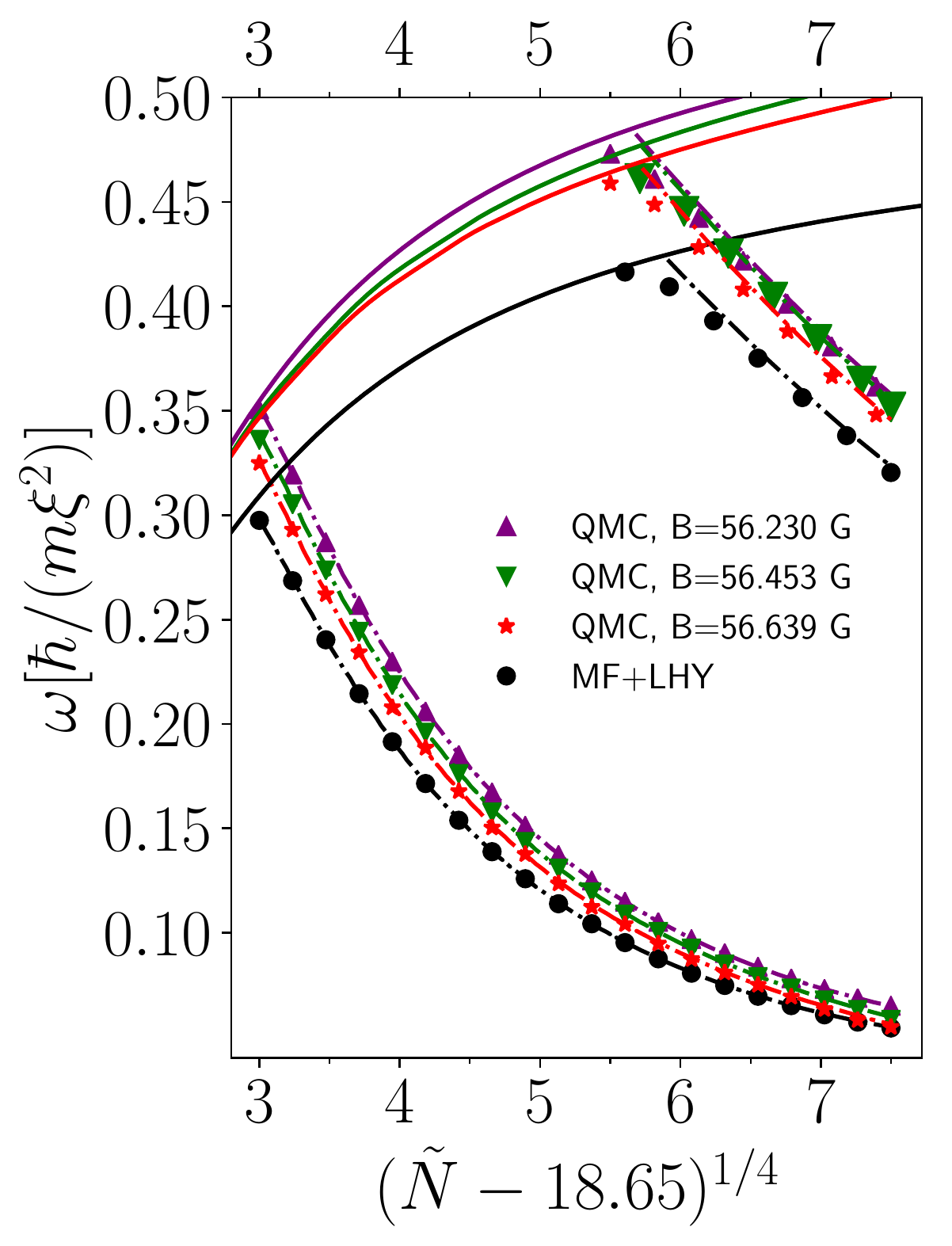}    
		\caption{Breathing (upper points) and quadrupole (lower points) frequencies as a function of the total atom number in  units of $\tilde{N}$. 
			Points are the results obtained from QMC and MF+LHY TDDFT calculations, and dashed lines
			are the $E_1/\hbar$ frequencies from the sum-rule approach (Eqs. (\ref{eq:monopole_E1}) and (\ref{eq:quadropole_E1})). Full lines represent the frequency corresponding to the absolute value of the droplet chemical potential $| \mu|$, corresponding to the legend from top to bottom
		}
		\label{fig:breathingmodecompareqmcmflhydiffb}
	\end{figure}

	We show in Fig. \ref{fig:breathingmodecompareqmcmflhydiffb}
	the breathing and quadrupole frequencies, corresponding to the more intense peaks, as a function of the number of atoms obtained with the QMC
	functional and the MF+LHY approach. For the latter, our results are in full agreement with those reported by Petrov using the
	Bovoliubov-de Gennes method \cite{petrov2015quantum}, which is fully equivalent to ours.
	The results are plotted in the universal units of the MF+LHY theory. 
	We find that the QMC functional predicts systematically larger monopole and quadrupole frequencies in all the range of particle numbers we have studied. 
	Additionally, as we change the magnetic field, i.e. the  scattering parameters, QMC predictions do not fall on the same curve, 
	meaning that the QMC functional breaks the MF+LHY universality.
	
	When the multipole strength is concentrated in a single narrow peak, it is possible to estimate the peak frequency using the sum rules approach
	\cite{bohigas1979sum,pitaevskii2016bose}. 
	Sum rules are energy moments of the strength function that, for some excitation operators,  can be written as 
	compact expressions involving expectation values on the ground state configuration. For the multipole operators considered here,
	two such sum rules are the linear-energy $m_1$ and cubic-energy $m_3$ sum rules. The inverse-energy sum rule $m_{-1}$ can be obtained from a 
	constrained calculation involving the Hamiltonian ${\cal H}'$ of Eq. (\ref{eq16}). Once determined, these three sum rules may be
	used to define two average energies $E_1 =\sqrt{m_1/m_{-1}}$  and  $E_3 =\sqrt{m_3/m_1}$ expecting, {\it bona fide}, that they are good estimates of
	the peak energy.
	
	For the monopole and quadrupole modes, the $E_1$ energies are \cite{bohigas1979sum}         
	\begin{equation}
	\label{eq:monopole_E1}
	E_1 (\ell=0)= 
	\sqrt{- \dfrac{4 \hbar^2}{m} \,      \dfrac{\ave{r^2}}{\left( \partial \ave{Q_0} / \partial \lambda \right) \rvert_{\lambda = 0}}} 
	\end{equation}
	and
	\begin{equation}
	\label{eq:quadropole_E1}
	E_1 (\ell=2) = \sqrt{-\dfrac{8 \hbar^2}{m} \, \dfrac{\ave{r^2}}{\left( \partial \ave{Q_2} / \partial \lambda  \right) \rvert_{\lambda = 0}}} \; ,
	\end{equation}
	with $\lambda$ being the parameter in the constrained Hamiltonian ${\cal H}'$, Eq.(\ref{eq16}),
	and $\ave{r^2} = \displaystyle \int d\vec{r}  \rho(r) r^2 / N$ evaluated at $\lambda = 0$.
	The frequencies corresponding to these energies are drawn in Figs. \ref{fig:monopole} and \ref{fig:quadrupole} as vertical dash-dotted lines.
	Except for small droplets, for which the monopole strength is very fragmented, one
	can see that they are good  estimates of the peak frequency.
	
	Closed expressions for the $E_3$ averages can be easiliy obtained for the monopole and the quadrupole modes \cite{bohigas1979sum,pitaevskii2016bose}. For the sake of completeness, we present the result obtained for the QMC functional. 
	
	Defining 
	\begin{eqnarray}
	E_{\alpha} &=& \alpha \int  d\mathbf{r} \rho^2(\mathbf{r})  
	\nonumber
	\\
	E_{\beta} &=& \beta \int  d\mathbf{r} \rho^{\gamma +1}(\mathbf{r})
	\nonumber
	\\
	\langle T \rangle & =& \frac{\hbar^2}{2m} \int d {\mathbf r} |\nabla \Psi({\mathbf r})|^2  \; ,
	\label{eq222}    
	\end{eqnarray}
	where $\Psi({\mathbf r})$  and $\rho(\mathbf{r})$  are those of the equilibrium configuration, we have 
	\begin{equation}
	\label{eq:monopole_E3}
	E_3 (\ell=0) = \left[\frac{\hbar^2}{N m \langle r^2 \rangle}\right]^{1/2} [4  \langle T \rangle + 9 (E_{\alpha} +  \gamma^2 E_{\beta})]^{1/2}
	\end{equation}
	\begin{equation}
	\label{eq:quadropole_E3}
	E_3 (\ell=2) = \left[\frac{\hbar^2}{N m \langle r^2 \rangle}\right]^{1/2} [4  \langle T \rangle]^{1/2}\; .
	\end{equation}
	We have $E_3 (\ell=2) < E_3 (\ell=0)$. The $\omega_3=E_3/\hbar$ frequencies are shown in Figs. \ref{fig:monopole} and  \ref{fig:quadrupole}
	as vertical dotted lines. It can be seen 
	that even when the strength is concentrated in a single peak, $\omega_3$ is a worse estimate of the peak frequency than  $\omega_1=E_1/\hbar$.
	This is likely so because $m_3$ gets contributions from the high energy part of the spectrum. At variance, since contributions to $m_{-1}$ 
	mainly come from the low energy part of the spectrum, $\omega_1$ is better suited for estimating the  peak frequency.
	
\begin{figure}
		\centering
		\includegraphics[width=\linewidth]{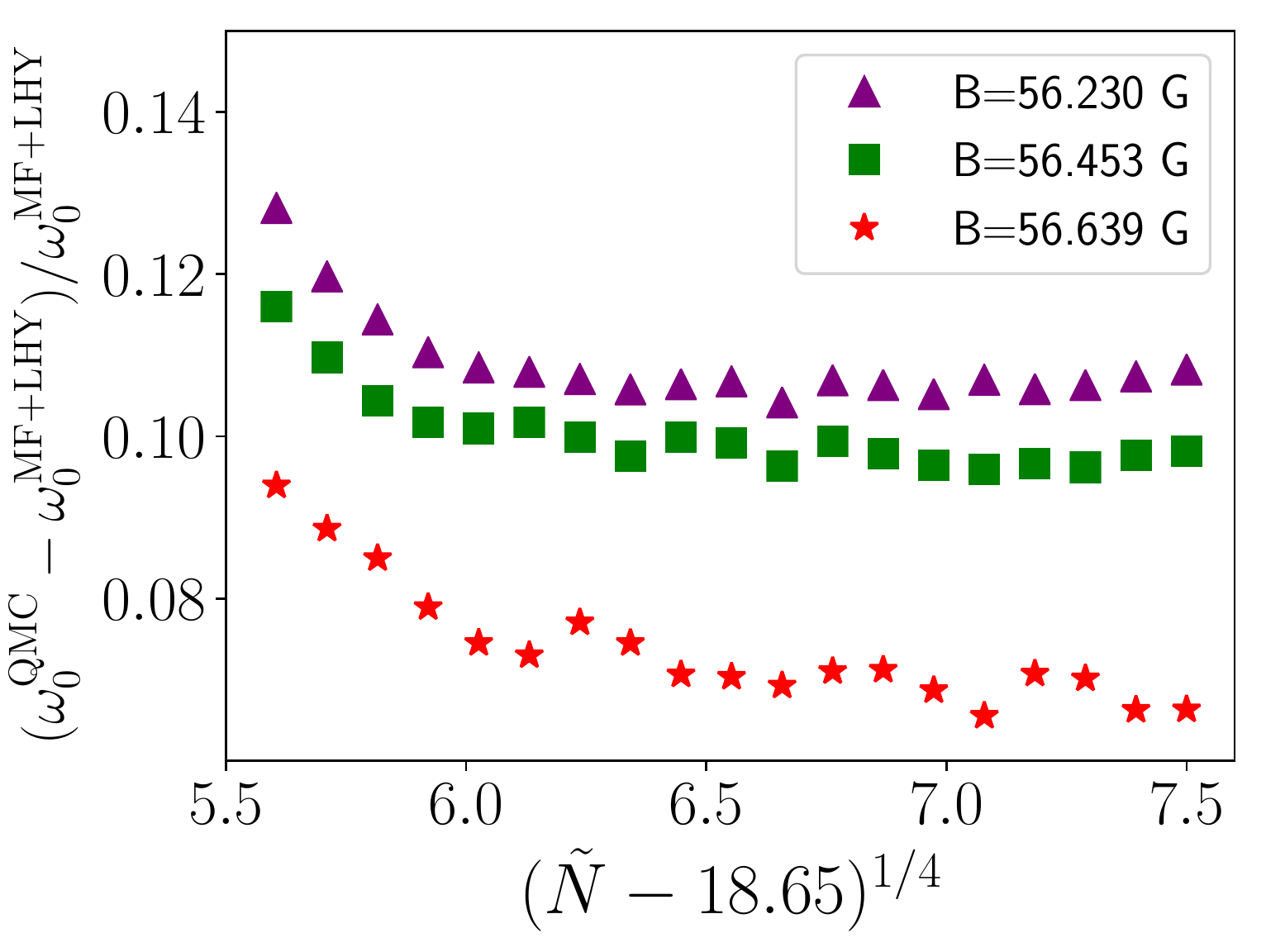}
		\includegraphics[width=\linewidth]{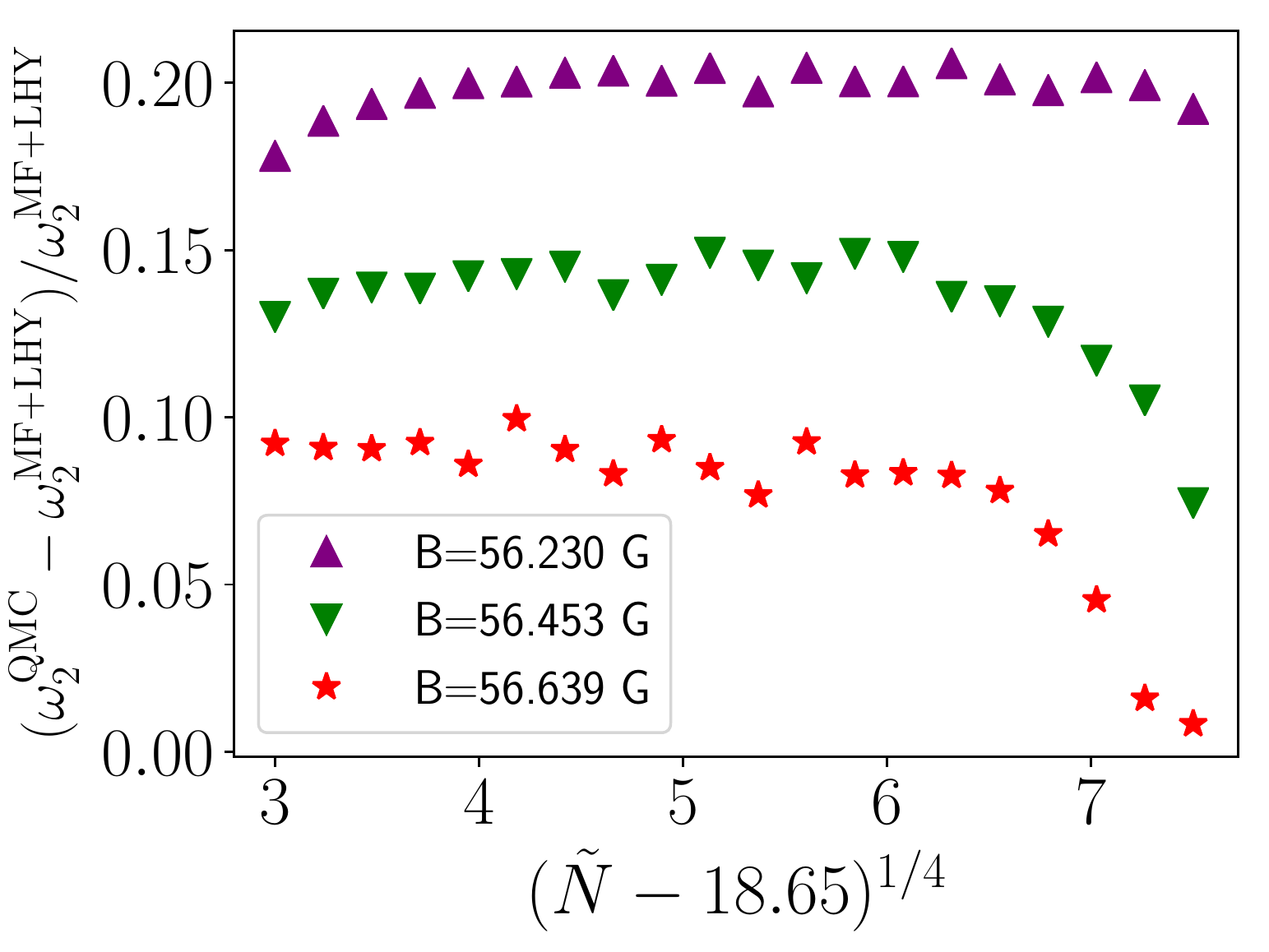}
		\caption{Relative frequency difference  between QMC and MF+LHY  TDDFT calculations for quadrupole (bottom figure) and monopole
			modes (top figure) as a function of the total atom number in  units of $\tilde{N}$.}
		\label{fig:relative_diff}
	\end{figure}

	The relative differences between the MF+LHY theory and the QMC functional for
	monopole and quadrupole frequencies  are presented in Fig. \ref{fig:relative_diff}. As the magnetic field 
	increases, the droplet is more correlated and differences of even $20\%$ can be observed.

\begin{figure}
		\centering
		\includegraphics[width=\linewidth]{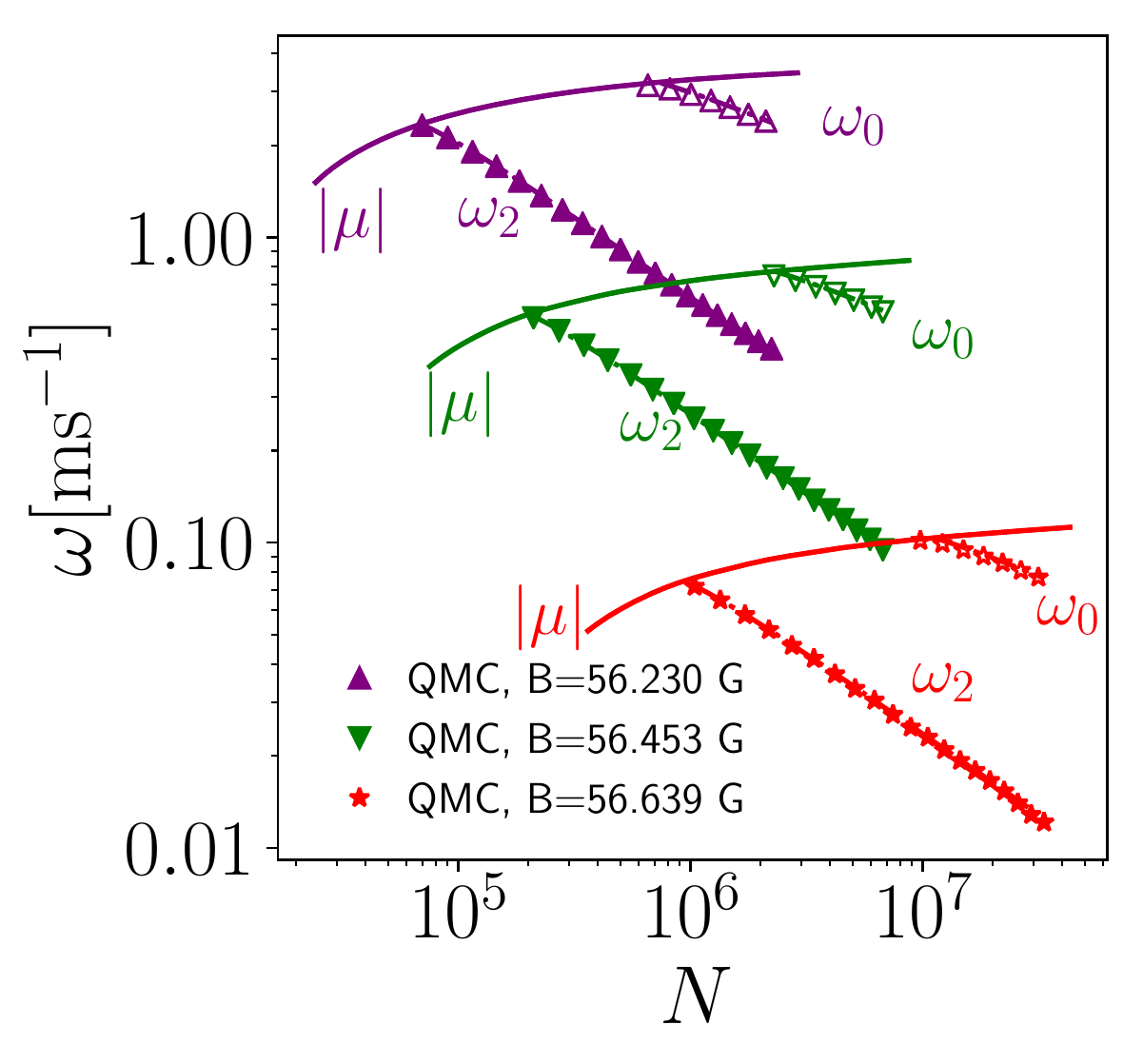}
		\caption{Predictions of the frequency corresponding to the absolute value of the chemical potential $|\mu|$, breathing frequency $\omega_0$ and quadropole frequency $\omega_2$ as a function of total atom number, using the QMC functionals. Dashed lines are $\omega_1  = E_1 / \hbar$ frequencies.}
		\label{fig:bothmodescompareqmcmflhydiffbrealisticunits}
	\end{figure}

	We finally compare in more detail the frequencies obtained with the QMC and MF+LHY functionals at  $B = 56.230$ G 
	for $\tilde{N} = 100$ and $\tilde{N} =1010$, which correspond to $N = 7 \times 10^4$ and $N= 7.1\times 10^5$, 
	respectively. Although it might require rather large droplets to observe neat breathing oscillations, systems with $\tilde{N}>100$, for which clean quadrupole modes show up (see Fig. \ref{fig:breathingmodecompareqmcmflhydiffb}),
	are already accessible in experiments \cite{cabrera2018quantum,semeghini2018self,ferioli2019collisions,derrico2019observation}. 
	For $N=7 \times 10^4$,   the quadrupole frequencies are
	$\omega_2^{\rm QMC} = 2323 \,{\rm Hz}$ and $\omega_2^{\rm MF+LHY} = 1972 \, {\rm Hz}$, i.e. oscillation 
	periods $\tau_2^{\rm QMC} = 2.70 \, {\rm ms}$ and $\tau_2^{\rm MF+LHY} = 3.19 \, {\rm ms}$. A similar comparison can 
	be made for the monopole frequency; for $N = 7.1 \times 10^5$,  the frequencies are
	$\omega_0^{\rm QMC} = 3114 \,{\rm Hz}$ and $\omega_0^{\rm MF+LHY} = 2755 \,{\rm Hz}$, and the 
	oscillation periods are $\tau_0^{\rm QMC} = 2.02 \,{\rm ms}$, and $\tau_0^{\rm MF+LHY} = 2.28 \,{\rm ms}$. In Fig. (\ref{fig:bothmodescompareqmcmflhydiffbrealisticunits}), we report our results for the breathing and quadrupole modes in not-reduced units to facilitate future comparisons with experiments.

	\section{\label{sec:summary} Summary and outlook}
	
	Using a new QMC-based density functional which properly incorporates
	finite-range effects, we have determined the monopole and quadrupole excitation modes of $^{39}$K quantum droplets at 
	the optimal MF+LHY mixture composition.  Comparing with the results obtained within the MF+LHY approach, we have found that finite-range effects have a 
	detectable influence on the excitation spectrum, whose study  may thus be a promising way to explore physics beyond the LHY correction.
	
	We have shown  that introducing the QMC functional into the usual DFT methodology can easily be done, as only  minor changes need to be made in the (many)  existing Gross-Pitaevskii numerical 
	solvers~\cite{antoine2015gpelab,wittek2015extended,schloss2018gpue}.
        This opens the door to using better functionals --based on including  quantum  effects beyond mean-field--	in the current applications of the extended Gross-Pitaevskii approach  \cite{astrakharchik2018dynamics,ferioli2019collisions}.
	
	The significant difference between the  predictions of QMC and MF+LHY functionals for the excitation spectrum 
	indicates that finite-range effects could show up 
	in other dynamical problems as well. In particular, in  droplet-droplet collisions \cite{ferioli2019collisions}, 
	where the actual value of the incompressibility might play a relevant role. A reliable functional might also be useful to
	study quantum droplet aspects that are currently  under study for superfluid $^4$He droplets, 
	as the appearance of quantum turbulence and of bulk and surface vorticity
	in  droplets merging; the equilibrium phase diagram of rotating quantum  droplets 
	\cite{ancilotto2018spinning,escartin2019vorticity,oconnell2020angular}, and the
	merging of vortex-hosting quantum droplets.  These aspects are at present under investigation.
	Further improvements in the building of a more accurate QMC functional should  consider the inclusion of surface tension effects others
	that those arising from the quantum kinetic energy term  \cite{marin2005free}. 
	
	\begin{acknowledgments}
			
	This work has been supported by the Ministerio de Economia, Industria y Competitividad (MINECO, Spain) under grants 
	Nos. FIS2017-84114-C2-1-P and FIS2017-87801-P (AEI/FEDER, UE), and by the EC Research Innovation Action under the H2020 Programme, 
	Project HPC-EUROPA3 (INFRAIA-2016-1-730897). V. C. gratefully acknowledges the support of G. E. Astrakharchik at the UPC and the 
	computer resources and technical support provided by Barcelona Supercomputing Center. We acknowledge financial support from Secretaria d'Universitats i Recerca del Departament d'Empresa i Coneixement de la Generalitat de Catalunya, co-funded by the European Union Regional Development Fund within the ERDF Operational Program of Catalunya (project QuantumCat, ref. 001-P-001644).
	
	\end{acknowledgments}

	\bibliography{references/references.bib} 

\begin{thebibliography}{55}%
\makeatletter
\providecommand \@ifxundefined [1]{%
 \@ifx{#1\undefined}
}%
\providecommand \@ifnum [1]{%
 \ifnum #1\expandafter \@firstoftwo
 \else \expandafter \@secondoftwo
 \fi
}%
\providecommand \@ifx [1]{%
 \ifx #1\expandafter \@firstoftwo
 \else \expandafter \@secondoftwo
 \fi
}%
\providecommand \natexlab [1]{#1}%
\providecommand \enquote  [1]{``#1''}%
\providecommand \bibnamefont  [1]{#1}%
\providecommand \bibfnamefont [1]{#1}%
\providecommand \citenamefont [1]{#1}%
\providecommand \href@noop [0]{\@secondoftwo}%
\providecommand \href [0]{\begingroup \@sanitize@url \@href}%
\providecommand \@href[1]{\@@startlink{#1}\@@href}%
\providecommand \@@href[1]{\endgroup#1\@@endlink}%
\providecommand \@sanitize@url [0]{\catcode `\\12\catcode `\$12\catcode
  `\&12\catcode `\#12\catcode `\^12\catcode `\_12\catcode `\%12\relax}%
\providecommand \@@startlink[1]{}%
\providecommand \@@endlink[0]{}%
\providecommand \url  [0]{\begingroup\@sanitize@url \@url }%
\providecommand \@url [1]{\endgroup\@href {#1}{\urlprefix }}%
\providecommand \urlprefix  [0]{URL }%
\providecommand \Eprint [0]{\href }%
\providecommand \doibase [0]{http://dx.doi.org/}%
\providecommand \selectlanguage [0]{\@gobble}%
\providecommand \bibinfo  [0]{\@secondoftwo}%
\providecommand \bibfield  [0]{\@secondoftwo}%
\providecommand \translation [1]{[#1]}%
\providecommand \BibitemOpen [0]{}%
\providecommand \bibitemStop [0]{}%
\providecommand \bibitemNoStop [0]{.\EOS\space}%
\providecommand \EOS [0]{\spacefactor3000\relax}%
\providecommand \BibitemShut  [1]{\csname bibitem#1\endcsname}%
\let\auto@bib@innerbib\@empty
\bibitem [{\citenamefont {Bloch}\ \emph {et~al.}(2008)\citenamefont {Bloch},
  \citenamefont {Dalibard},\ and\ \citenamefont {Zwerger}}]{bloch2008many}%
  \BibitemOpen
  \bibfield  {author} {\bibinfo {author} {\bibfnamefont {I.}~\bibnamefont
  {Bloch}}, \bibinfo {author} {\bibfnamefont {J.}~\bibnamefont {Dalibard}}, \
  and\ \bibinfo {author} {\bibfnamefont {W.}~\bibnamefont {Zwerger}},\ }\href
  {\doibase 10.1103/RevModPhys.80.885} {\bibfield  {journal} {\bibinfo
  {journal} {Rev. Mod. Phys.}\ }\textbf {\bibinfo {volume} {80}},\ \bibinfo
  {pages} {885} (\bibinfo {year} {2008})}\BibitemShut {NoStop}%
\bibitem [{\citenamefont {Pethick}\ and\ \citenamefont
  {Smith}(2008)}]{pethick2008bose}%
  \BibitemOpen
  \bibfield  {author} {\bibinfo {author} {\bibfnamefont {C.~J.}\ \bibnamefont
  {Pethick}}\ and\ \bibinfo {author} {\bibfnamefont {H.}~\bibnamefont
  {Smith}},\ }\href {\doibase 10.1017/CBO9780511802850} {\emph {\bibinfo
  {title} {Bose-Einstein condensation in dilute gases}}}\ (\bibinfo
  {publisher} {Cambridge University Press},\ \bibinfo {year}
  {2008})\BibitemShut {NoStop}%
\bibitem [{\citenamefont {Pitaevskii}\ and\ \citenamefont
  {Stringari}(2016)}]{pitaevskii2016bose}%
  \BibitemOpen
  \bibfield  {author} {\bibinfo {author} {\bibfnamefont {L.}~\bibnamefont
  {Pitaevskii}}\ and\ \bibinfo {author} {\bibfnamefont {S.}~\bibnamefont
  {Stringari}},\ }\href {\doibase 10.1093/acprof:oso/9780198758884.001.0001}
  {\emph {\bibinfo {title} {Bose-Einstein condensation and superfluidity}}},\
  Vol.\ \bibinfo {volume} {164}\ (\bibinfo  {publisher} {Oxford University
  Press},\ \bibinfo {year} {2016})\BibitemShut {NoStop}%
\bibitem [{\citenamefont {Bogoliubov}(1947)}]{bogoliubov1947theory}%
  \BibitemOpen
  \bibfield  {author} {\bibinfo {author} {\bibfnamefont {N.}~\bibnamefont
  {Bogoliubov}},\ }\href@noop {} {\bibfield  {journal} {\bibinfo  {journal} {J.
  Phys. (USSR)}\ }\textbf {\bibinfo {volume} {11}},\ \bibinfo {pages} {23}
  (\bibinfo {year} {1947})}\BibitemShut {NoStop}%
\bibitem [{\citenamefont {Cikojevi\'{c}}\ \emph {et~al.}(2018)\citenamefont
  {Cikojevi\'{c}}, \citenamefont {D\v{z}elalija}, \citenamefont
  {Stipanovi\'{c}}, \citenamefont {Vranje\v{s}~Marki\'{c}},\ and\ \citenamefont
  {Boronat}}]{cikojevic2018ultradilute}%
  \BibitemOpen
  \bibfield  {author} {\bibinfo {author} {\bibfnamefont {V.}~\bibnamefont
  {Cikojevi\'{c}}}, \bibinfo {author} {\bibfnamefont {K.}~\bibnamefont
  {D\v{z}elalija}}, \bibinfo {author} {\bibfnamefont {P.}~\bibnamefont
  {Stipanovi\'{c}}}, \bibinfo {author} {\bibfnamefont {L.}~\bibnamefont
  {Vranje\v{s}~Marki\'{c}}}, \ and\ \bibinfo {author} {\bibfnamefont
  {J.}~\bibnamefont {Boronat}},\ }\href {\doibase 10.1103/PhysRevB.97.140502}
  {\bibfield  {journal} {\bibinfo  {journal} {Phys. Rev. B}\ }\textbf {\bibinfo
  {volume} {97}},\ \bibinfo {pages} {140502(R)} (\bibinfo {year}
  {2018})}\BibitemShut {NoStop}%
\bibitem [{\citenamefont {Cikojevi{\'{c}}}\ \emph {et~al.}(2019)\citenamefont
  {Cikojevi{\'{c}}}, \citenamefont {Marki{\'{c}}}, \citenamefont
  {Astrakharchik},\ and\ \citenamefont {Boronat}}]{cikojevic2019universality}%
  \BibitemOpen
  \bibfield  {author} {\bibinfo {author} {\bibfnamefont {V.}~\bibnamefont
  {Cikojevi{\'{c}}}}, \bibinfo {author} {\bibfnamefont {L.~V.}\ \bibnamefont
  {Marki{\'{c}}}}, \bibinfo {author} {\bibfnamefont {G.~E.}\ \bibnamefont
  {Astrakharchik}}, \ and\ \bibinfo {author} {\bibfnamefont {J.}~\bibnamefont
  {Boronat}},\ }\href {\doibase 10.1103/PhysRevA.99.023618} {\bibfield
  {journal} {\bibinfo  {journal} {Phys. Rev. A}\ }\textbf {\bibinfo {volume}
  {99}},\ \bibinfo {pages} {023618} (\bibinfo {year} {2019})}\BibitemShut
  {NoStop}%
\bibitem [{\citenamefont {Parisi}\ and\ \citenamefont
  {Giorgini}(2020)}]{parisi2020quantum}%
  \BibitemOpen
  \bibfield  {author} {\bibinfo {author} {\bibfnamefont {L.}~\bibnamefont
  {Parisi}}\ and\ \bibinfo {author} {\bibfnamefont {S.}~\bibnamefont
  {Giorgini}},\ }\href {https://arxiv.org/abs/2003.05231} {\bibfield  {journal}
  {\bibinfo  {journal} {arXiv preprint arXiv:2003.05231}\ } (\bibinfo {year}
  {2020})}\BibitemShut {NoStop}%
\bibitem [{\citenamefont {Parisi}\ \emph {et~al.}(2019)\citenamefont {Parisi},
  \citenamefont {Astrakharchik},\ and\ \citenamefont
  {Giorgini}}]{parisi2019liquid}%
  \BibitemOpen
  \bibfield  {author} {\bibinfo {author} {\bibfnamefont {L.}~\bibnamefont
  {Parisi}}, \bibinfo {author} {\bibfnamefont {G.~E.}\ \bibnamefont
  {Astrakharchik}}, \ and\ \bibinfo {author} {\bibfnamefont {S.}~\bibnamefont
  {Giorgini}},\ }\href {\doibase 10.1103/PhysRevLett.122.105302} {\bibfield
  {journal} {\bibinfo  {journal} {Phys. Rev. Lett}\ }\textbf {\bibinfo {volume}
  {122}},\ \bibinfo {pages} {105302} (\bibinfo {year} {2019})}\BibitemShut
  {NoStop}%
\bibitem [{\citenamefont {Staudinger}\ \emph {et~al.}(2018)\citenamefont
  {Staudinger}, \citenamefont {Mazzanti},\ and\ \citenamefont
  {Zillich}}]{staudinger2018self}%
  \BibitemOpen
  \bibfield  {author} {\bibinfo {author} {\bibfnamefont {C.}~\bibnamefont
  {Staudinger}}, \bibinfo {author} {\bibfnamefont {F.}~\bibnamefont
  {Mazzanti}}, \ and\ \bibinfo {author} {\bibfnamefont {R.~E.}\ \bibnamefont
  {Zillich}},\ }\href {\doibase 10.1103/PhysRevA.98.023633} {\bibfield
  {journal} {\bibinfo  {journal} {Phys. Rev. A}\ }\textbf {\bibinfo {volume}
  {98}},\ \bibinfo {pages} {023633} (\bibinfo {year} {2018})}\BibitemShut
  {NoStop}%
\bibitem [{\citenamefont {Petrov}\ and\ \citenamefont
  {Astrakharchik}(2016)}]{petrov2016ultradilute}%
  \BibitemOpen
  \bibfield  {author} {\bibinfo {author} {\bibfnamefont {D.~S.}\ \bibnamefont
  {Petrov}}\ and\ \bibinfo {author} {\bibfnamefont {G.~E.}\ \bibnamefont
  {Astrakharchik}},\ }\href {\doibase 10.1103/PhysRevLett.117.100401}
  {\bibfield  {journal} {\bibinfo  {journal} {Phys. Rev. Lett}\ }\textbf
  {\bibinfo {volume} {117}},\ \bibinfo {pages} {100401} (\bibinfo {year}
  {2016})}\BibitemShut {NoStop}%
\bibitem [{\citenamefont {Bombin}\ \emph {et~al.}(2017)\citenamefont {Bombin},
  \citenamefont {Boronat},\ and\ \citenamefont {Mazzanti}}]{bombin2017dipolar}%
  \BibitemOpen
  \bibfield  {author} {\bibinfo {author} {\bibfnamefont {R.}~\bibnamefont
  {Bombin}}, \bibinfo {author} {\bibfnamefont {J.}~\bibnamefont {Boronat}}, \
  and\ \bibinfo {author} {\bibfnamefont {F.}~\bibnamefont {Mazzanti}},\ }\href
  {\doibase 10.1103/PhysRevLett.119.250402} {\bibfield  {journal} {\bibinfo
  {journal} {Phys. Rev. Lett}\ }\textbf {\bibinfo {volume} {119}},\ \bibinfo
  {pages} {250402} (\bibinfo {year} {2017})}\BibitemShut {NoStop}%
\bibitem [{\citenamefont {Ancilotto}\ \emph
  {et~al.}(2018{\natexlab{a}})\citenamefont {Ancilotto}, \citenamefont
  {Barranco}, \citenamefont {Guilleumas},\ and\ \citenamefont
  {Pi}}]{ancilotto2018self}%
  \BibitemOpen
  \bibfield  {author} {\bibinfo {author} {\bibfnamefont {F.}~\bibnamefont
  {Ancilotto}}, \bibinfo {author} {\bibfnamefont {M.}~\bibnamefont {Barranco}},
  \bibinfo {author} {\bibfnamefont {M.}~\bibnamefont {Guilleumas}}, \ and\
  \bibinfo {author} {\bibfnamefont {M.}~\bibnamefont {Pi}},\ }\href {\doibase
  10.1103/PhysRevA.98.053623} {\bibfield  {journal} {\bibinfo  {journal} {Phys.
  Rev. A}\ }\textbf {\bibinfo {volume} {98}},\ \bibinfo {pages} {053623}
  (\bibinfo {year} {2018}{\natexlab{a}})}\BibitemShut {NoStop}%
\bibitem [{\citenamefont {Hu}\ and\ \citenamefont
  {Liu}(2020{\natexlab{a}})}]{hu2020consistent}%
  \BibitemOpen
  \bibfield  {author} {\bibinfo {author} {\bibfnamefont {H.}~\bibnamefont
  {Hu}}\ and\ \bibinfo {author} {\bibfnamefont {X.-J.}\ \bibnamefont {Liu}},\
  }\href {https://arxiv.org/abs/2005.08581v1} {\bibfield  {journal} {\bibinfo
  {journal} {arXiv preprint arXiv:2005.08581}\ } (\bibinfo {year}
  {2020}{\natexlab{a}})}\BibitemShut {NoStop}%
\bibitem [{\citenamefont {Hu}\ and\ \citenamefont
  {Liu}(2020{\natexlab{b}})}]{hu2020microscopic}%
  \BibitemOpen
  \bibfield  {author} {\bibinfo {author} {\bibfnamefont {H.}~\bibnamefont
  {Hu}}\ and\ \bibinfo {author} {\bibfnamefont {X.-J.}\ \bibnamefont {Liu}},\
  }\href {https://arxiv.org/abs/2006.00434v1} {\bibfield  {journal} {\bibinfo
  {journal} {arXiv preprint arXiv:2006.00434}\ } (\bibinfo {year}
  {2020}{\natexlab{b}})}\BibitemShut {NoStop}%
\bibitem [{\citenamefont {Ota}\ and\ \citenamefont
  {Astrakharchik}(2020)}]{ota2020beyond}%
  \BibitemOpen
  \bibfield  {author} {\bibinfo {author} {\bibfnamefont {M.}~\bibnamefont
  {Ota}}\ and\ \bibinfo {author} {\bibfnamefont {G.~E.}\ \bibnamefont
  {Astrakharchik}},\ }\href {https://arxiv.org/abs/2005.10047v1} {\bibfield
  {journal} {\bibinfo  {journal} {arXiv preprint arXiv:2005.10047}\ } (\bibinfo
  {year} {2020})}\BibitemShut {NoStop}%
\bibitem [{\citenamefont {Petrov}(2015)}]{petrov2015quantum}%
  \BibitemOpen
  \bibfield  {author} {\bibinfo {author} {\bibfnamefont {D.~S.}\ \bibnamefont
  {Petrov}},\ }\href {\doibase 10.1103/PhysRevLett.115.155302} {\bibfield
  {journal} {\bibinfo  {journal} {Phys. Rev. Lett}\ }\textbf {\bibinfo {volume}
  {115}},\ \bibinfo {pages} {155302} (\bibinfo {year} {2015})}\BibitemShut
  {NoStop}%
\bibitem [{\citenamefont {Lee}\ \emph {et~al.}(1957)\citenamefont {Lee},
  \citenamefont {Huang},\ and\ \citenamefont {Yang}}]{lee1957eigenvalues}%
  \BibitemOpen
  \bibfield  {author} {\bibinfo {author} {\bibfnamefont {T.~D.}\ \bibnamefont
  {Lee}}, \bibinfo {author} {\bibfnamefont {K.}~\bibnamefont {Huang}}, \ and\
  \bibinfo {author} {\bibfnamefont {C.~N.}\ \bibnamefont {Yang}},\ }\href
  {\doibase 10.1103/PhysRev.106.1135} {\bibfield  {journal} {\bibinfo
  {journal} {Phys. Rev.}\ }\textbf {\bibinfo {volume} {106}},\ \bibinfo {pages}
  {1135} (\bibinfo {year} {1957})}\BibitemShut {NoStop}%
\bibitem [{\citenamefont {Barranco}\ \emph {et~al.}(2006)\citenamefont
  {Barranco}, \citenamefont {Guardiola}, \citenamefont {Hern\'andez},
  \citenamefont {Mayol}, \citenamefont {Navarro},\ and\ \citenamefont
  {Pi}}]{barranco2006helium}%
  \BibitemOpen
  \bibfield  {author} {\bibinfo {author} {\bibfnamefont {M.}~\bibnamefont
  {Barranco}}, \bibinfo {author} {\bibfnamefont {R.}~\bibnamefont {Guardiola}},
  \bibinfo {author} {\bibfnamefont {S.}~\bibnamefont {Hern\'andez}}, \bibinfo
  {author} {\bibfnamefont {R.}~\bibnamefont {Mayol}}, \bibinfo {author}
  {\bibfnamefont {J.}~\bibnamefont {Navarro}}, \ and\ \bibinfo {author}
  {\bibfnamefont {M.}~\bibnamefont {Pi}},\ }\href
  {https://doi.org/10.1007/s10909-005-9267-0} {\bibfield  {journal} {\bibinfo
  {journal} {J. Low Temp. Phys.}\ }\textbf {\bibinfo {volume} {142}},\ \bibinfo
  {pages} {1} (\bibinfo {year} {2006})}\BibitemShut {NoStop}%
\bibitem [{\citenamefont {Cabrera}\ \emph {et~al.}(2018)\citenamefont
  {Cabrera}, \citenamefont {Tanzi}, \citenamefont {Sanz}, \citenamefont
  {Naylor}, \citenamefont {Thomas}, \citenamefont {Cheiney},\ and\
  \citenamefont {Tarruell}}]{cabrera2018quantum}%
  \BibitemOpen
  \bibfield  {author} {\bibinfo {author} {\bibfnamefont {C.}~\bibnamefont
  {Cabrera}}, \bibinfo {author} {\bibfnamefont {L.}~\bibnamefont {Tanzi}},
  \bibinfo {author} {\bibfnamefont {J.}~\bibnamefont {Sanz}}, \bibinfo {author}
  {\bibfnamefont {B.}~\bibnamefont {Naylor}}, \bibinfo {author} {\bibfnamefont
  {P.}~\bibnamefont {Thomas}}, \bibinfo {author} {\bibfnamefont
  {P.}~\bibnamefont {Cheiney}}, \ and\ \bibinfo {author} {\bibfnamefont
  {L.}~\bibnamefont {Tarruell}},\ }\href {10.1126/science.aao5686} {\bibfield
  {journal} {\bibinfo  {journal} {Science}\ }\textbf {\bibinfo {volume}
  {359}},\ \bibinfo {pages} {301} (\bibinfo {year} {2018})}\BibitemShut
  {NoStop}%
\bibitem [{\citenamefont {Semeghini}\ \emph {et~al.}(2018)\citenamefont
  {Semeghini}, \citenamefont {Ferioli}, \citenamefont {Masi}, \citenamefont
  {Mazzinghi}, \citenamefont {Wolswijk}, \citenamefont {Minardi}, \citenamefont
  {Modugno}, \citenamefont {Modugno}, \citenamefont {Inguscio},\ and\
  \citenamefont {Fattori}}]{semeghini2018self}%
  \BibitemOpen
  \bibfield  {author} {\bibinfo {author} {\bibfnamefont {G.}~\bibnamefont
  {Semeghini}}, \bibinfo {author} {\bibfnamefont {G.}~\bibnamefont {Ferioli}},
  \bibinfo {author} {\bibfnamefont {L.}~\bibnamefont {Masi}}, \bibinfo {author}
  {\bibfnamefont {C.}~\bibnamefont {Mazzinghi}}, \bibinfo {author}
  {\bibfnamefont {L.}~\bibnamefont {Wolswijk}}, \bibinfo {author}
  {\bibfnamefont {F.}~\bibnamefont {Minardi}}, \bibinfo {author} {\bibfnamefont
  {M.}~\bibnamefont {Modugno}}, \bibinfo {author} {\bibfnamefont
  {G.}~\bibnamefont {Modugno}}, \bibinfo {author} {\bibfnamefont
  {M.}~\bibnamefont {Inguscio}}, \ and\ \bibinfo {author} {\bibfnamefont
  {M.}~\bibnamefont {Fattori}},\ }\href {\doibase
  10.1103/PhysRevLett.120.235301} {\bibfield  {journal} {\bibinfo  {journal}
  {Phys. Rev. Lett}\ }\textbf {\bibinfo {volume} {120}},\ \bibinfo {pages}
  {235301} (\bibinfo {year} {2018})}\BibitemShut {NoStop}%
\bibitem [{\citenamefont {Cheiney}\ \emph {et~al.}(2018)\citenamefont
  {Cheiney}, \citenamefont {Cabrera}, \citenamefont {Sanz}, \citenamefont
  {Naylor}, \citenamefont {Tanzi},\ and\ \citenamefont
  {Tarruell}}]{cheiney2018bright}%
  \BibitemOpen
  \bibfield  {author} {\bibinfo {author} {\bibfnamefont {P.}~\bibnamefont
  {Cheiney}}, \bibinfo {author} {\bibfnamefont {C.~R.}\ \bibnamefont
  {Cabrera}}, \bibinfo {author} {\bibfnamefont {J.}~\bibnamefont {Sanz}},
  \bibinfo {author} {\bibfnamefont {B.}~\bibnamefont {Naylor}}, \bibinfo
  {author} {\bibfnamefont {L.}~\bibnamefont {Tanzi}}, \ and\ \bibinfo {author}
  {\bibfnamefont {L.}~\bibnamefont {Tarruell}},\ }\href {\doibase
  10.1103/PhysRevLett.120.135301} {\bibfield  {journal} {\bibinfo  {journal}
  {Phys. Rev. Lett}\ }\textbf {\bibinfo {volume} {120}},\ \bibinfo {pages}
  {135301} (\bibinfo {year} {2018})}\BibitemShut {NoStop}%
\bibitem [{\citenamefont {D'Errico}\ \emph {et~al.}(2019)\citenamefont
  {D'Errico}, \citenamefont {Burchianti}, \citenamefont {Prevedelli},
  \citenamefont {Salasnich}, \citenamefont {Ancilotto}, \citenamefont
  {Modugno}, \citenamefont {Minardi},\ and\ \citenamefont
  {Fort}}]{derrico2019observation}%
  \BibitemOpen
  \bibfield  {author} {\bibinfo {author} {\bibfnamefont {C.}~\bibnamefont
  {D'Errico}}, \bibinfo {author} {\bibfnamefont {A.}~\bibnamefont
  {Burchianti}}, \bibinfo {author} {\bibfnamefont {M.}~\bibnamefont
  {Prevedelli}}, \bibinfo {author} {\bibfnamefont {L.}~\bibnamefont
  {Salasnich}}, \bibinfo {author} {\bibfnamefont {F.}~\bibnamefont
  {Ancilotto}}, \bibinfo {author} {\bibfnamefont {M.}~\bibnamefont {Modugno}},
  \bibinfo {author} {\bibfnamefont {F.}~\bibnamefont {Minardi}}, \ and\
  \bibinfo {author} {\bibfnamefont {C.}~\bibnamefont {Fort}},\ }\href {\doibase
  10.1103/PhysRevResearch.1.033155} {\bibfield  {journal} {\bibinfo  {journal}
  {Phys. Rev. Research}\ }\textbf {\bibinfo {volume} {1}},\ \bibinfo {pages}
  {033155} (\bibinfo {year} {2019})}\BibitemShut {NoStop}%
\bibitem [{\citenamefont {Cikojevi{\'{c}}}\ \emph {et~al.}(2020)\citenamefont
  {Cikojevi{\'{c}}}, \citenamefont {Marki{\'{c}}},\ and\ \citenamefont
  {Boronat}}]{cikojevic2020finite}%
  \BibitemOpen
  \bibfield  {author} {\bibinfo {author} {\bibfnamefont {V.}~\bibnamefont
  {Cikojevi{\'{c}}}}, \bibinfo {author} {\bibfnamefont {L.~V.}\ \bibnamefont
  {Marki{\'{c}}}}, \ and\ \bibinfo {author} {\bibfnamefont {J.}~\bibnamefont
  {Boronat}},\ }\href {\doibase 10.1088/1367-2630/ab867a} {\bibfield  {journal}
  {\bibinfo  {journal} {New J. Phys}\ }\textbf {\bibinfo {volume} {22}},\
  \bibinfo {pages} {053045} (\bibinfo {year} {2020})}\BibitemShut {NoStop}%
\bibitem [{\citenamefont {Tanzi}\ \emph {et~al.}(2018)\citenamefont {Tanzi},
  \citenamefont {Cabrera}, \citenamefont {Sanz}, \citenamefont {Cheiney},
  \citenamefont {Tomza},\ and\ \citenamefont {Tarruell}}]{tanzi2018feshbach}%
  \BibitemOpen
  \bibfield  {author} {\bibinfo {author} {\bibfnamefont {L.}~\bibnamefont
  {Tanzi}}, \bibinfo {author} {\bibfnamefont {C.~R.}\ \bibnamefont {Cabrera}},
  \bibinfo {author} {\bibfnamefont {J.}~\bibnamefont {Sanz}}, \bibinfo {author}
  {\bibfnamefont {P.}~\bibnamefont {Cheiney}}, \bibinfo {author} {\bibfnamefont
  {M.}~\bibnamefont {Tomza}}, \ and\ \bibinfo {author} {\bibfnamefont
  {L.}~\bibnamefont {Tarruell}},\ }\href {\doibase 10.1103/PhysRevA.98.062712}
  {\bibfield  {journal} {\bibinfo  {journal} {Phys. Rev. A}\ }\textbf {\bibinfo
  {volume} {98}},\ \bibinfo {pages} {062712} (\bibinfo {year}
  {2018})}\BibitemShut {NoStop}%
\bibitem [{\citenamefont {Newton}(2013)}]{newton2013scattering}%
  \BibitemOpen
  \bibfield  {author} {\bibinfo {author} {\bibfnamefont {R.~G.}\ \bibnamefont
  {Newton}},\ }\href {\doibase 10.1007/978-3-642-88128-2} {\emph {\bibinfo
  {title} {Scattering Theory of Waves and Particles}}}\ (\bibinfo  {publisher}
  {Springer Science \& Business Media},\ \bibinfo {year} {2013})\BibitemShut
  {NoStop}%
\bibitem [{\citenamefont {Tononi}(2019)}]{tononi2019zero}%
  \BibitemOpen
  \bibfield  {author} {\bibinfo {author} {\bibfnamefont {A.}~\bibnamefont
  {Tononi}},\ }\href {\doibase https://doi.org/10.3390/condmat4010020}
  {\bibfield  {journal} {\bibinfo  {journal} {Condens. Matter}\ }\textbf
  {\bibinfo {volume} {4}},\ \bibinfo {pages} {20} (\bibinfo {year}
  {2019})}\BibitemShut {NoStop}%
\bibitem [{\citenamefont {Tononi}\ \emph {et~al.}(2018)\citenamefont {Tononi},
  \citenamefont {Cappellaro},\ and\ \citenamefont
  {Salasnich}}]{tononi2018condensation}%
  \BibitemOpen
  \bibfield  {author} {\bibinfo {author} {\bibfnamefont {A.}~\bibnamefont
  {Tononi}}, \bibinfo {author} {\bibfnamefont {A.}~\bibnamefont {Cappellaro}},
  \ and\ \bibinfo {author} {\bibfnamefont {L.}~\bibnamefont {Salasnich}},\
  }\href {\doibase 10.1088/1367-2630/aaf75e} {\bibfield  {journal} {\bibinfo
  {journal} {New J. Phys}\ }\textbf {\bibinfo {volume} {20}},\ \bibinfo {pages}
  {125007} (\bibinfo {year} {2018})}\BibitemShut {NoStop}%
\bibitem [{\citenamefont {Salasnich}(2017)}]{salasnich2017nonuniversal}%
  \BibitemOpen
  \bibfield  {author} {\bibinfo {author} {\bibfnamefont {L.}~\bibnamefont
  {Salasnich}},\ }\href {\doibase 10.1103/PhysRevLett.118.130402} {\bibfield
  {journal} {\bibinfo  {journal} {Phys. Rev. Lett}\ }\textbf {\bibinfo {volume}
  {118}},\ \bibinfo {pages} {130402} (\bibinfo {year} {2017})}\BibitemShut
  {NoStop}%
\bibitem [{\citenamefont {Boronat}\ and\ \citenamefont
  {Casulleras}(1994)}]{boronat1994monte}%
  \BibitemOpen
  \bibfield  {author} {\bibinfo {author} {\bibfnamefont {J.}~\bibnamefont
  {Boronat}}\ and\ \bibinfo {author} {\bibfnamefont {J.}~\bibnamefont
  {Casulleras}},\ }\href {\doibase 10.1103/PhysRevB.49.8920} {\bibfield
  {journal} {\bibinfo  {journal} {Phys. Rev. B}\ }\textbf {\bibinfo {volume}
  {49}},\ \bibinfo {pages} {8920} (\bibinfo {year} {1994})}\BibitemShut
  {NoStop}%
\bibitem [{\citenamefont {Giorgini}\ \emph {et~al.}(1999)\citenamefont
  {Giorgini}, \citenamefont {Boronat},\ and\ \citenamefont
  {Casulleras}}]{giorgini1999ground}%
  \BibitemOpen
  \bibfield  {author} {\bibinfo {author} {\bibfnamefont {S.}~\bibnamefont
  {Giorgini}}, \bibinfo {author} {\bibfnamefont {J.}~\bibnamefont {Boronat}}, \
  and\ \bibinfo {author} {\bibfnamefont {J.}~\bibnamefont {Casulleras}},\
  }\href {\doibase 10.1103/PhysRevA.60.5129} {\bibfield  {journal} {\bibinfo
  {journal} {Phys. Rev. A}\ }\textbf {\bibinfo {volume} {60}},\ \bibinfo
  {pages} {5129} (\bibinfo {year} {1999})}\BibitemShut {NoStop}%
\bibitem [{\citenamefont {J\o{}rgensen}\ \emph {et~al.}(2018)\citenamefont
  {J\o{}rgensen}, \citenamefont {Bruun},\ and\ \citenamefont
  {Arlt}}]{jorgensen2018dilute}%
  \BibitemOpen
  \bibfield  {author} {\bibinfo {author} {\bibfnamefont {N.~B.}\ \bibnamefont
  {J\o{}rgensen}}, \bibinfo {author} {\bibfnamefont {G.~M.}\ \bibnamefont
  {Bruun}}, \ and\ \bibinfo {author} {\bibfnamefont {J.~J.}\ \bibnamefont
  {Arlt}},\ }\href {\doibase 10.1103/PhysRevLett.121.173403} {\bibfield
  {journal} {\bibinfo  {journal} {Phys. Rev. Lett}\ }\textbf {\bibinfo {volume}
  {121}},\ \bibinfo {pages} {173403} (\bibinfo {year} {2018})}\BibitemShut
  {NoStop}%
\bibitem [{\citenamefont {Roy}\ \emph {et~al.}(2013)\citenamefont {Roy},
  \citenamefont {Landini}, \citenamefont {Trenkwalder}, \citenamefont
  {Semeghini}, \citenamefont {Spagnolli}, \citenamefont {Simoni}, \citenamefont
  {Fattori}, \citenamefont {Inguscio},\ and\ \citenamefont
  {Modugno}}]{roy2013test}%
  \BibitemOpen
  \bibfield  {author} {\bibinfo {author} {\bibfnamefont {S.}~\bibnamefont
  {Roy}}, \bibinfo {author} {\bibfnamefont {M.}~\bibnamefont {Landini}},
  \bibinfo {author} {\bibfnamefont {A.}~\bibnamefont {Trenkwalder}}, \bibinfo
  {author} {\bibfnamefont {G.}~\bibnamefont {Semeghini}}, \bibinfo {author}
  {\bibfnamefont {G.}~\bibnamefont {Spagnolli}}, \bibinfo {author}
  {\bibfnamefont {A.}~\bibnamefont {Simoni}}, \bibinfo {author} {\bibfnamefont
  {M.}~\bibnamefont {Fattori}}, \bibinfo {author} {\bibfnamefont
  {M.}~\bibnamefont {Inguscio}}, \ and\ \bibinfo {author} {\bibfnamefont
  {G.}~\bibnamefont {Modugno}},\ }\href {\doibase
  10.1103/PhysRevLett.111.053202} {\bibfield  {journal} {\bibinfo  {journal}
  {Phys. Rev. Lett}\ }\textbf {\bibinfo {volume} {111}},\ \bibinfo {pages}
  {053202} (\bibinfo {year} {2013})}\BibitemShut {NoStop}%
\bibitem [{\citenamefont {Ferioli}\ \emph {et~al.}(2019)\citenamefont
  {Ferioli}, \citenamefont {Semeghini}, \citenamefont {Masi}, \citenamefont
  {Giusti}, \citenamefont {Modugno}, \citenamefont {Inguscio}, \citenamefont
  {Gallem\'{\i}}, \citenamefont {Recati},\ and\ \citenamefont
  {Fattori}}]{ferioli2019collisions}%
  \BibitemOpen
  \bibfield  {author} {\bibinfo {author} {\bibfnamefont {G.}~\bibnamefont
  {Ferioli}}, \bibinfo {author} {\bibfnamefont {G.}~\bibnamefont {Semeghini}},
  \bibinfo {author} {\bibfnamefont {L.}~\bibnamefont {Masi}}, \bibinfo {author}
  {\bibfnamefont {G.}~\bibnamefont {Giusti}}, \bibinfo {author} {\bibfnamefont
  {G.}~\bibnamefont {Modugno}}, \bibinfo {author} {\bibfnamefont
  {M.}~\bibnamefont {Inguscio}}, \bibinfo {author} {\bibfnamefont
  {A.}~\bibnamefont {Gallem\'{\i}}}, \bibinfo {author} {\bibfnamefont
  {A.}~\bibnamefont {Recati}}, \ and\ \bibinfo {author} {\bibfnamefont
  {M.}~\bibnamefont {Fattori}},\ }\href {\doibase
  10.1103/PhysRevLett.122.090401} {\bibfield  {journal} {\bibinfo  {journal}
  {Phys. Rev. Lett}\ }\textbf {\bibinfo {volume} {122}},\ \bibinfo {pages}
  {090401} (\bibinfo {year} {2019})}\BibitemShut {NoStop}%
\bibitem [{\citenamefont {Stringari}\ and\ \citenamefont
  {Treiner}(1987)}]{stringari1985surface}%
  \BibitemOpen
  \bibfield  {author} {\bibinfo {author} {\bibfnamefont {S.}~\bibnamefont
  {Stringari}}\ and\ \bibinfo {author} {\bibfnamefont {J.}~\bibnamefont
  {Treiner}},\ }\href {\doibase 10.1103/PhysRevB.36.8369} {\bibfield  {journal}
  {\bibinfo  {journal} {Phys. Rev. B}\ }\textbf {\bibinfo {volume} {36}},\
  \bibinfo {pages} {8369} (\bibinfo {year} {1987})}\BibitemShut {NoStop}%
\bibitem [{\citenamefont {Ancilotto}\ \emph {et~al.}(2017)\citenamefont
  {Ancilotto}, \citenamefont {Barranco}, \citenamefont {Coppens}, \citenamefont
  {Eloranta}, \citenamefont {Halberstadt}, \citenamefont {Hernando},
  \citenamefont {Mateo},\ and\ \citenamefont {Pi}}]{ancilotto2017density}%
  \BibitemOpen
  \bibfield  {author} {\bibinfo {author} {\bibfnamefont {F.}~\bibnamefont
  {Ancilotto}}, \bibinfo {author} {\bibfnamefont {M.}~\bibnamefont {Barranco}},
  \bibinfo {author} {\bibfnamefont {F.}~\bibnamefont {Coppens}}, \bibinfo
  {author} {\bibfnamefont {J.}~\bibnamefont {Eloranta}}, \bibinfo {author}
  {\bibfnamefont {N.}~\bibnamefont {Halberstadt}}, \bibinfo {author}
  {\bibfnamefont {A.}~\bibnamefont {Hernando}}, \bibinfo {author}
  {\bibfnamefont {D.}~\bibnamefont {Mateo}}, \ and\ \bibinfo {author}
  {\bibfnamefont {M.}~\bibnamefont {Pi}},\ }\href {\doibase
  10.1080/0144235X.2017.1351672} {\bibfield  {journal} {\bibinfo  {journal}
  {Int. Rev. Phys. Chem.}\ }\textbf {\bibinfo {volume} {36}},\ \bibinfo {pages}
  {621} (\bibinfo {year} {2017})},\ \Eprint
  {http://arxiv.org/abs/https://doi.org/10.1080/0144235X.2017.1351672}
  {https://doi.org/10.1080/0144235X.2017.1351672} \BibitemShut {NoStop}%
\bibitem [{\citenamefont {Chin}\ \emph {et~al.}(2009)\citenamefont {Chin},
  \citenamefont {Janecek},\ and\ \citenamefont {Krotscheck}}]{chin2009any}%
  \BibitemOpen
  \bibfield  {author} {\bibinfo {author} {\bibfnamefont {S.~A.}\ \bibnamefont
  {Chin}}, \bibinfo {author} {\bibfnamefont {S.}~\bibnamefont {Janecek}}, \
  and\ \bibinfo {author} {\bibfnamefont {E.}~\bibnamefont {Krotscheck}},\
  }\href {\doibase https://doi.org/10.1016/j.cplett.2009.01.068} {\bibfield
  {journal} {\bibinfo  {journal} {Chemical Phys. Let.}\ }\textbf {\bibinfo
  {volume} {470}},\ \bibinfo {pages} {342 } (\bibinfo {year}
  {2009})}\BibitemShut {NoStop}%
\bibitem [{\citenamefont {Dalfovo}\ \emph {et~al.}(1999)\citenamefont
  {Dalfovo}, \citenamefont {Giorgini}, \citenamefont {Pitaevskii},\ and\
  \citenamefont {Stringari}}]{dalfovo1999theory}%
  \BibitemOpen
  \bibfield  {author} {\bibinfo {author} {\bibfnamefont {F.}~\bibnamefont
  {Dalfovo}}, \bibinfo {author} {\bibfnamefont {S.}~\bibnamefont {Giorgini}},
  \bibinfo {author} {\bibfnamefont {L.~P.}\ \bibnamefont {Pitaevskii}}, \ and\
  \bibinfo {author} {\bibfnamefont {S.}~\bibnamefont {Stringari}},\ }\href
  {\doibase 10.1103/RevModPhys.71.463} {\bibfield  {journal} {\bibinfo
  {journal} {Rev. Mod. Phys.}\ }\textbf {\bibinfo {volume} {71}},\ \bibinfo
  {pages} {463} (\bibinfo {year} {1999})}\BibitemShut {NoStop}%
\bibitem [{\citenamefont {Baillie}\ \emph {et~al.}(2017)\citenamefont
  {Baillie}, \citenamefont {Wilson},\ and\ \citenamefont
  {Blakie}}]{baillie2017collective}%
  \BibitemOpen
  \bibfield  {author} {\bibinfo {author} {\bibfnamefont {D.}~\bibnamefont
  {Baillie}}, \bibinfo {author} {\bibfnamefont {R.~M.}\ \bibnamefont {Wilson}},
  \ and\ \bibinfo {author} {\bibfnamefont {P.~B.}\ \bibnamefont {Blakie}},\
  }\href {\doibase 10.1103/PhysRevLett.119.255302} {\bibfield  {journal}
  {\bibinfo  {journal} {Phys. Rev. Lett}\ }\textbf {\bibinfo {volume} {119}},\
  \bibinfo {pages} {255302} (\bibinfo {year} {2017})}\BibitemShut {NoStop}%
\bibitem [{\citenamefont {Stringari}\ and\ \citenamefont
  {Vautherin}(1979)}]{stringari1979damping}%
  \BibitemOpen
  \bibfield  {author} {\bibinfo {author} {\bibfnamefont {S.}~\bibnamefont
  {Stringari}}\ and\ \bibinfo {author} {\bibfnamefont {D.}~\bibnamefont
  {Vautherin}},\ }\href
  {https://www.sciencedirect.com/science/journal/03702693/88/1} {\bibfield
  {journal} {\bibinfo  {journal} {Phys. Let. B}\ }\textbf {\bibinfo {volume}
  {88}},\ \bibinfo {pages} {1} (\bibinfo {year} {1979})}\BibitemShut {NoStop}%
\bibitem [{\citenamefont {Pi}\ \emph {et~al.}(1986)\citenamefont {Pi},
  \citenamefont {Barranco}, \citenamefont {Nemeth}, \citenamefont {Ngô},\ and\
  \citenamefont {Tomasi}}]{pi1986time}%
  \BibitemOpen
  \bibfield  {author} {\bibinfo {author} {\bibfnamefont {M.}~\bibnamefont
  {Pi}}, \bibinfo {author} {\bibfnamefont {M.}~\bibnamefont {Barranco}},
  \bibinfo {author} {\bibfnamefont {J.}~\bibnamefont {Nemeth}}, \bibinfo
  {author} {\bibfnamefont {C.}~\bibnamefont {Ngô}}, \ and\ \bibinfo {author}
  {\bibfnamefont {E.}~\bibnamefont {Tomasi}},\ }\href {\doibase
  https://doi.org/10.1016/0370-2693(86)91143-3} {\bibfield  {journal} {\bibinfo
   {journal} {Phys. Let. B}\ }\textbf {\bibinfo {volume} {166}},\ \bibinfo
  {pages} {1 } (\bibinfo {year} {1986})}\BibitemShut {NoStop}%
\bibitem [{\citenamefont {Jin}\ \emph {et~al.}(1996)\citenamefont {Jin},
  \citenamefont {Ensher}, \citenamefont {Matthews}, \citenamefont {Wieman},\
  and\ \citenamefont {Cornell}}]{jin1996collective}%
  \BibitemOpen
  \bibfield  {author} {\bibinfo {author} {\bibfnamefont {D.~S.}\ \bibnamefont
  {Jin}}, \bibinfo {author} {\bibfnamefont {J.~R.}\ \bibnamefont {Ensher}},
  \bibinfo {author} {\bibfnamefont {M.~R.}\ \bibnamefont {Matthews}}, \bibinfo
  {author} {\bibfnamefont {C.~E.}\ \bibnamefont {Wieman}}, \ and\ \bibinfo
  {author} {\bibfnamefont {E.~A.}\ \bibnamefont {Cornell}},\ }\href {\doibase
  10.1103/PhysRevLett.77.420} {\bibfield  {journal} {\bibinfo  {journal} {Phys.
  Rev. Lett}\ }\textbf {\bibinfo {volume} {77}},\ \bibinfo {pages} {420}
  (\bibinfo {year} {1996})}\BibitemShut {NoStop}%
\bibitem [{\citenamefont {Altmeyer}\ \emph {et~al.}(2007)\citenamefont
  {Altmeyer}, \citenamefont {Riedl}, \citenamefont {Kohstall}, \citenamefont
  {Wright}, \citenamefont {Geursen}, \citenamefont {Bartenstein}, \citenamefont
  {Chin}, \citenamefont {Denschlag},\ and\ \citenamefont
  {Grimm}}]{altmeyer2007precision}%
  \BibitemOpen
  \bibfield  {author} {\bibinfo {author} {\bibfnamefont {A.}~\bibnamefont
  {Altmeyer}}, \bibinfo {author} {\bibfnamefont {S.}~\bibnamefont {Riedl}},
  \bibinfo {author} {\bibfnamefont {C.}~\bibnamefont {Kohstall}}, \bibinfo
  {author} {\bibfnamefont {M.~J.}\ \bibnamefont {Wright}}, \bibinfo {author}
  {\bibfnamefont {R.}~\bibnamefont {Geursen}}, \bibinfo {author} {\bibfnamefont
  {M.}~\bibnamefont {Bartenstein}}, \bibinfo {author} {\bibfnamefont
  {C.}~\bibnamefont {Chin}}, \bibinfo {author} {\bibfnamefont {J.~H.}\
  \bibnamefont {Denschlag}}, \ and\ \bibinfo {author} {\bibfnamefont
  {R.}~\bibnamefont {Grimm}},\ }\href {\doibase 10.1103/PhysRevLett.98.040401}
  {\bibfield  {journal} {\bibinfo  {journal} {Phys. Rev. Lett}\ }\textbf
  {\bibinfo {volume} {98}},\ \bibinfo {pages} {040401} (\bibinfo {year}
  {2007})}\BibitemShut {NoStop}%
\bibitem [{\citenamefont {Bohigas}\ \emph {et~al.}(1979)\citenamefont
  {Bohigas}, \citenamefont {Lane},\ and\ \citenamefont
  {Martorell}}]{bohigas1979sum}%
  \BibitemOpen
  \bibfield  {author} {\bibinfo {author} {\bibfnamefont {O.}~\bibnamefont
  {Bohigas}}, \bibinfo {author} {\bibfnamefont {A.}~\bibnamefont {Lane}}, \
  and\ \bibinfo {author} {\bibfnamefont {J.}~\bibnamefont {Martorell}},\ }\href
  {\doibase https://doi.org/10.1016/0370-1573(79)90079-6} {\bibfield  {journal}
  {\bibinfo  {journal} {Phys. Rep.}\ }\textbf {\bibinfo {volume} {51}},\
  \bibinfo {pages} {267 } (\bibinfo {year} {1979})}\BibitemShut {NoStop}%
\bibitem [{\citenamefont {Ferioli}\ \emph {et~al.}(2020)\citenamefont
  {Ferioli}, \citenamefont {Semeghini}, \citenamefont {Terradas-Brians\'o},
  \citenamefont {Masi}, \citenamefont {Fattori},\ and\ \citenamefont
  {Modugno}}]{ferioli2020dynamical}%
  \BibitemOpen
  \bibfield  {author} {\bibinfo {author} {\bibfnamefont {G.}~\bibnamefont
  {Ferioli}}, \bibinfo {author} {\bibfnamefont {G.}~\bibnamefont {Semeghini}},
  \bibinfo {author} {\bibfnamefont {S.}~\bibnamefont {Terradas-Brians\'o}},
  \bibinfo {author} {\bibfnamefont {L.}~\bibnamefont {Masi}}, \bibinfo {author}
  {\bibfnamefont {M.}~\bibnamefont {Fattori}}, \ and\ \bibinfo {author}
  {\bibfnamefont {M.}~\bibnamefont {Modugno}},\ }\href {\doibase
  10.1103/PhysRevResearch.2.013269} {\bibfield  {journal} {\bibinfo  {journal}
  {Phys. Rev. Research}\ }\textbf {\bibinfo {volume} {2}},\ \bibinfo {pages}
  {013269} (\bibinfo {year} {2020})}\BibitemShut {NoStop}%
\bibitem [{\citenamefont {Serra}\ \emph {et~al.}(1991)\citenamefont {Serra},
  \citenamefont {Navarro}, \citenamefont {Barranco},\ and\ \citenamefont
  {Van~Giai}}]{serra1991collective}%
  \BibitemOpen
  \bibfield  {author} {\bibinfo {author} {\bibfnamefont {L.}~\bibnamefont
  {Serra}}, \bibinfo {author} {\bibfnamefont {J.}~\bibnamefont {Navarro}},
  \bibinfo {author} {\bibfnamefont {M.}~\bibnamefont {Barranco}}, \ and\
  \bibinfo {author} {\bibfnamefont {N.}~\bibnamefont {Van~Giai}},\ }\href
  {\doibase 10.1103/PhysRevLett.67.2311} {\bibfield  {journal} {\bibinfo
  {journal} {Phys. Rev. Lett}\ }\textbf {\bibinfo {volume} {67}},\ \bibinfo
  {pages} {2311} (\bibinfo {year} {1991})}\BibitemShut {NoStop}%
\bibitem [{\citenamefont {Barranco}\ and\ \citenamefont
  {Hern\'andez}(1994)}]{barranco1994response}%
  \BibitemOpen
  \bibfield  {author} {\bibinfo {author} {\bibfnamefont {M.}~\bibnamefont
  {Barranco}}\ and\ \bibinfo {author} {\bibfnamefont {E.~S.}\ \bibnamefont
  {Hern\'andez}},\ }\href {\doibase 10.1103/PhysRevB.49.12078} {\bibfield
  {journal} {\bibinfo  {journal} {Phys. Rev. B}\ }\textbf {\bibinfo {volume}
  {49}},\ \bibinfo {pages} {12078} (\bibinfo {year} {1994})}\BibitemShut
  {NoStop}%
\bibitem [{\citenamefont {Br\"uhl}\ \emph {et~al.}(2004)\citenamefont
  {Br\"uhl}, \citenamefont {Guardiola}, \citenamefont {Kalinin}, \citenamefont
  {Kornilov}, \citenamefont {Navarro}, \citenamefont {Savas},\ and\
  \citenamefont {Toennies}}]{bruhl2004diffraction}%
  \BibitemOpen
  \bibfield  {author} {\bibinfo {author} {\bibfnamefont {R.}~\bibnamefont
  {Br\"uhl}}, \bibinfo {author} {\bibfnamefont {R.}~\bibnamefont {Guardiola}},
  \bibinfo {author} {\bibfnamefont {A.}~\bibnamefont {Kalinin}}, \bibinfo
  {author} {\bibfnamefont {O.}~\bibnamefont {Kornilov}}, \bibinfo {author}
  {\bibfnamefont {J.}~\bibnamefont {Navarro}}, \bibinfo {author} {\bibfnamefont
  {T.}~\bibnamefont {Savas}}, \ and\ \bibinfo {author} {\bibfnamefont {J.~P.}\
  \bibnamefont {Toennies}},\ }\href {\doibase 10.1103/PhysRevLett.92.185301}
  {\bibfield  {journal} {\bibinfo  {journal} {Phys. Rev. Lett}\ }\textbf
  {\bibinfo {volume} {92}},\ \bibinfo {pages} {185301} (\bibinfo {year}
  {2004})}\BibitemShut {NoStop}%
\bibitem [{\citenamefont {Antoine}\ and\ \citenamefont
  {Duboscq}(2015)}]{antoine2015gpelab}%
  \BibitemOpen
  \bibfield  {author} {\bibinfo {author} {\bibfnamefont {X.}~\bibnamefont
  {Antoine}}\ and\ \bibinfo {author} {\bibfnamefont {R.}~\bibnamefont
  {Duboscq}},\ }\href {\doibase https://doi.org/10.1016/j.cpc.2015.03.012}
  {\bibfield  {journal} {\bibinfo  {journal} {Comput. Phys. Commun}\ }\textbf
  {\bibinfo {volume} {193}},\ \bibinfo {pages} {95 } (\bibinfo {year}
  {2015})}\BibitemShut {NoStop}%
\bibitem [{\citenamefont {Wittek}\ and\ \citenamefont
  {Calderaro}(2015)}]{wittek2015extended}%
  \BibitemOpen
  \bibfield  {author} {\bibinfo {author} {\bibfnamefont {P.}~\bibnamefont
  {Wittek}}\ and\ \bibinfo {author} {\bibfnamefont {L.}~\bibnamefont
  {Calderaro}},\ }\href {\doibase https://doi.org/10.1016/j.cpc.2015.07.017}
  {\bibfield  {journal} {\bibinfo  {journal} {Comput. Phys. Commun}\ }\textbf
  {\bibinfo {volume} {197}},\ \bibinfo {pages} {339 } (\bibinfo {year}
  {2015})}\BibitemShut {NoStop}%
\bibitem [{\citenamefont {Schloss}\ and\ \citenamefont
  {O'Riordan}(2018)}]{schloss2018gpue}%
  \BibitemOpen
  \bibfield  {author} {\bibinfo {author} {\bibfnamefont {J.~R.}\ \bibnamefont
  {Schloss}}\ and\ \bibinfo {author} {\bibfnamefont {L.~J.}\ \bibnamefont
  {O'Riordan}},\ }\href {\doibase https://doi.org/10.21105/joss.01037}
  {\bibfield  {journal} {\bibinfo  {journal} {J. Open Source Softw}\ }\textbf
  {\bibinfo {volume} {3}},\ \bibinfo {pages} {1037} (\bibinfo {year}
  {2018})}\BibitemShut {NoStop}%
\bibitem [{\citenamefont {Astrakharchik}\ and\ \citenamefont
  {Malomed}(2018)}]{astrakharchik2018dynamics}%
  \BibitemOpen
  \bibfield  {author} {\bibinfo {author} {\bibfnamefont {G.~E.}\ \bibnamefont
  {Astrakharchik}}\ and\ \bibinfo {author} {\bibfnamefont {B.~A.}\ \bibnamefont
  {Malomed}},\ }\href {\doibase 10.1103/PhysRevA.98.013631} {\bibfield
  {journal} {\bibinfo  {journal} {Phys. Rev. A}\ }\textbf {\bibinfo {volume}
  {98}},\ \bibinfo {pages} {013631} (\bibinfo {year} {2018})}\BibitemShut
  {NoStop}%
\bibitem [{\citenamefont {Ancilotto}\ \emph
  {et~al.}(2018{\natexlab{b}})\citenamefont {Ancilotto}, \citenamefont
  {Barranco},\ and\ \citenamefont {Pi}}]{ancilotto2018spinning}%
  \BibitemOpen
  \bibfield  {author} {\bibinfo {author} {\bibfnamefont {F.}~\bibnamefont
  {Ancilotto}}, \bibinfo {author} {\bibfnamefont {M.}~\bibnamefont {Barranco}},
  \ and\ \bibinfo {author} {\bibfnamefont {M.}~\bibnamefont {Pi}},\ }\href
  {\doibase 10.1103/PhysRevB.97.184515} {\bibfield  {journal} {\bibinfo
  {journal} {Phys. Rev. B}\ }\textbf {\bibinfo {volume} {97}},\ \bibinfo
  {pages} {184515} (\bibinfo {year} {2018}{\natexlab{b}})}\BibitemShut
  {NoStop}%
\bibitem [{\citenamefont {Escart\'{\i}n}\ \emph {et~al.}(2019)\citenamefont
  {Escart\'{\i}n}, \citenamefont {Ancilotto}, \citenamefont {Barranco},\ and\
  \citenamefont {Pi}}]{escartin2019vorticity}%
  \BibitemOpen
  \bibfield  {author} {\bibinfo {author} {\bibfnamefont {J.~M.}\ \bibnamefont
  {Escart\'{\i}n}}, \bibinfo {author} {\bibfnamefont {F.}~\bibnamefont
  {Ancilotto}}, \bibinfo {author} {\bibfnamefont {M.}~\bibnamefont {Barranco}},
  \ and\ \bibinfo {author} {\bibfnamefont {M.}~\bibnamefont {Pi}},\ }\href
  {\doibase 10.1103/PhysRevB.99.140505} {\bibfield  {journal} {\bibinfo
  {journal} {Phys. Rev. B}\ }\textbf {\bibinfo {volume} {99}},\ \bibinfo
  {pages} {140505(R)} (\bibinfo {year} {2019})}\BibitemShut {NoStop}%
\bibitem [{\citenamefont {O'Connell}\ \emph {et~al.}(2020)\citenamefont
  {O'Connell}, \citenamefont {Tanyag}, \citenamefont {Verma}, \citenamefont
  {Bernando}, \citenamefont {Pang}, \citenamefont {Bacellar}, \citenamefont
  {Saladrigas}, \citenamefont {Mahl}, \citenamefont {Toulson}, \citenamefont
  {Kumagai}, \citenamefont {Walter}, \citenamefont {Ancilotto}, \citenamefont
  {Barranco}, \citenamefont {Pi}, \citenamefont {Bostedt}, \citenamefont
  {Gessner},\ and\ \citenamefont {Vilesov}}]{oconnell2020angular}%
  \BibitemOpen
  \bibfield  {author} {\bibinfo {author} {\bibfnamefont {S.~M.~O.}\
  \bibnamefont {O'Connell}}, \bibinfo {author} {\bibfnamefont {R.~M.~P.}\
  \bibnamefont {Tanyag}}, \bibinfo {author} {\bibfnamefont {D.}~\bibnamefont
  {Verma}}, \bibinfo {author} {\bibfnamefont {C.}~\bibnamefont {Bernando}},
  \bibinfo {author} {\bibfnamefont {W.}~\bibnamefont {Pang}}, \bibinfo {author}
  {\bibfnamefont {C.}~\bibnamefont {Bacellar}}, \bibinfo {author}
  {\bibfnamefont {C.~A.}\ \bibnamefont {Saladrigas}}, \bibinfo {author}
  {\bibfnamefont {J.}~\bibnamefont {Mahl}}, \bibinfo {author} {\bibfnamefont
  {B.~W.}\ \bibnamefont {Toulson}}, \bibinfo {author} {\bibfnamefont
  {Y.}~\bibnamefont {Kumagai}}, \bibinfo {author} {\bibfnamefont
  {P.}~\bibnamefont {Walter}}, \bibinfo {author} {\bibfnamefont
  {F.}~\bibnamefont {Ancilotto}}, \bibinfo {author} {\bibfnamefont
  {M.}~\bibnamefont {Barranco}}, \bibinfo {author} {\bibfnamefont
  {M.}~\bibnamefont {Pi}}, \bibinfo {author} {\bibfnamefont {C.}~\bibnamefont
  {Bostedt}}, \bibinfo {author} {\bibfnamefont {O.}~\bibnamefont {Gessner}}, \
  and\ \bibinfo {author} {\bibfnamefont {A.~F.}\ \bibnamefont {Vilesov}},\
  }\href {\doibase 10.1103/PhysRevLett.124.215301} {\bibfield  {journal}
  {\bibinfo  {journal} {Phys. Rev. Lett.}\ }\textbf {\bibinfo {volume} {124}},\
  \bibinfo {pages} {215301} (\bibinfo {year} {2020})}\BibitemShut {NoStop}%
\bibitem [{\citenamefont {Mar\'{\i}n}\ \emph {et~al.}(2005)\citenamefont
  {Mar\'{\i}n}, \citenamefont {Boronat},\ and\ \citenamefont
  {Casulleras}}]{marin2005free}%
  \BibitemOpen
  \bibfield  {author} {\bibinfo {author} {\bibfnamefont {J.~M.}\ \bibnamefont
  {Mar\'{\i}n}}, \bibinfo {author} {\bibfnamefont {J.}~\bibnamefont {Boronat}},
  \ and\ \bibinfo {author} {\bibfnamefont {J.}~\bibnamefont {Casulleras}},\
  }\href {\doibase 10.1103/PhysRevB.71.144518} {\bibfield  {journal} {\bibinfo
  {journal} {Phys. Rev. B}\ }\textbf {\bibinfo {volume} {71}},\ \bibinfo
  {pages} {144518} (\bibinfo {year} {2005})}\BibitemShut {NoStop}%
\end{thebibliography}%

\end{document}